\documentclass[iop,revtex4]{emulateapj}
\usepackage{amsmath,amssymb}
\usepackage{graphicx}
\usepackage{natbib}

\slugcomment{}

\shorttitle{Network Analysis of Cosmic Structures}
\shortauthors{Hong \& Dey}

\begin{document}

%define the title

%\title{Network Analysis on Cosmic Structures : Beyond Correlation Functions}
\title{Network Analysis of Cosmic Structures : Network Centrality and Topological Environment}
%in cosmological smoothed particle hydrodynamics simulations}

\author{ 
Sungryong Hong\altaffilmark{1} and  
Arjun Dey\altaffilmark{1}
}

\altaffiltext{1}{National Optical Astronomy Observatory,  
Tucson, AZ 85719, USA}

\begin{abstract}
We apply simple analyses techniques developed for the study of complex networks to the study of the cosmic web, 
the large scale galaxy distribution. 
In this paper, we measure three network centralities (ranks of topological importance), 
Degree Centrality (DC), Closeness Centrality (CL), and Betweenness Centrality (BC) from a network 
built from the Cosmological Evolution Survey (COSMOS) catalog. 
We define 8 galaxy populations according to the centrality measures; 
Void, Wall, and Cluster by DC, Main Branch and Dangling Leaf by BC, 
and Kernel, Backbone, and Fracture by CL. 
We also define three populations by voronoi tessellation density to compare these with the DC selection.
We apply the topological selections to galaxies in the (photometric) redshift range $0.91<z<0.94$ from the COSMOS survey, 
and explore whether the red and blue galaxy populations show differences in color, 
star-formation rate (SFR) and stellar mass in the different topological regions. 
Despite the limitations and uncertainties associated with using photometric redshift and indirect measurements 
of galactic parameters, the preliminary results illustrate the potential of network analysis. 
The coming future surveys will provide better statistical samples to test and improve this ``network cosmology''. 

\end{abstract}

%generate the tableofcontents automatically
%\clearpage
%\tableofcontents
%\clearpage

%\tableofcontents

%\clearpage
\section{Introduction}

Studies of galaxy evolution have now definitively established that the evolution of galaxies depends to some extent on their environment 
(e.g., Davis \& Geller 1976, Postman \& Geller 1984, Butcher \& Oemler 1984, Dressler et al. 1997, Balogh et al. 1999, McGee et al. 2011, Giodini et al. 2012, Dressler et al. 2013). Most of these studies have attempted to correlate observable properties of galaxies with a simple measure of the environmental density, 
usually derived from the local number density of galaxies, measured either by the counts in an aperture or by the distance 
to the $n^{th}$ nearest neighbor (e.g., Dressler 1980, Blanton et al. 2005, Cooper et al. 2005, Cooper et al. 2006, Prescott et al. 2008, Mayo et al. 2012, Scoville et al. 2013). 
However, density may not be the only environmental parameter driving galaxy evolution; it is conceivable that the local {\it topology} of the matter distribution plays an important role in galaxy evolution by affecting matter accretion, merging rates, and the efficacy of feedback. 

The large scale matter distribution of the Universe has rich geometric and topological features. Numerical simulations of increasing sophistication have demonstrated that this large scale structure is formed through cosmic time by gravitational instabilities that originate in the initially almost featureless gaussian random field that characterizes the matter distribution in the early universe (Davis et al. 1985, Springel et al. 2005, and Vogelsberger et al. 2014). While we can not observe the full matter distribution directly, we can trace it by the spatial distribution of galaxies. Numerous imaging and spectroscopic surveys of the sky have revealed this complex structure 
(e.g., de Lapparent, Geller, \& Huchra 1986, Adams et al. 2011, Dawson et al 2013). 
Galaxies are arrayed in a filamentary distribution (commonly referred to as the ``cosmic web''; e.g., Bond et al. 1996, Colless et al. 2003, Tegmark et al. 2004, Huchra et al. 2005) that intersects at dense clusters and bounds voids. In order to understand the evolution of galaxies in different structures, we first need robust ways of characterizing the topology. 

To characterize the large scale structure of the cosmic web, various methods have been adopted from other fields of science. Correlation functions of the galaxy point distribution, pioneered by Peebles (1980), have been long used to understand the galaxy distribution. The 2-point correlation function (and the related power spectrum) is a powerful measure of the clustering strength of a given galaxy population and its use has demonstrated that different galaxy populations exhibit different correlation strengths (e.g., Landy \& Szalay 1993, Padmanabhan et al. 2007). The higher order correlation functions, while containing valuable information on the higher order moments of the distribution (i.e., the topology), require very large galaxy samples and become increasingly computationally expensive (Sheth \& Bhuvnesh 2003, Budav\'ari et al. 2003).  Genus numbers have also been used to characterize the overall topology of the galaxy distribution (Gott, Weinberg, \& Melott 1987, Choi et al. 2010). Several methodologies have been employed to identify specific filamentary structures in the galaxy distribution: minimum spanning trees (Barrow et al. 1985); the ``Candy'' model (Stoica et al. 2005); wavelets (Martinez et al. 2005); Hessian matrices; and Minkowski functionals of the density field (Sheth et al. 2003, Arag\'{o}n-Calvo et al. 2007, Sousbie et al. 2008, Bond et al. 2010).

% For example, to identify filamentary structures, Zeldovich et al.  (1982) introduced percolation analysis, while Barrow et al. (1985) tested a very different idea of the minimum spanning tree of a graph representation of galaxy distribution.  More recently, Stoica et al (2005) utilized the Candy model and Martinez et al. (2005) exploited the wavelet analysis to detect filamentary structures.  To cover more generic morphological features, the Hessian Matrices and Minkowski Functionals of density field have been analyzed by many authors (Sheth et al. 2003, Arag\'{o}n-Calvo et al. 2007, Sousbie et al. 2008, Bond et al. 2010 ).

This wide spectrum of applied methodologies reflects how difficult to characterize cosmic structures in a single robust framework. In this paper, we attempt a different approach to identifying topological features in the large scale galaxy distribution, drawing from the field of network science. Network science is a branch of graph theory focused on identifying the key interrelationships and topologies within complex networks (e.g., Barabasi 2009, Newman 2010). With roots in Euler's classic solution to the K\"onigsburg bridge problem (Euler 1741), network science was mainly used in the last century to analyze social networks. However, during the last two decades it has experienced a rapid growth in analyses tools, tools and understanding, driven largely by the growth of the Internet, the World Wide Web and computing power. Here, as a first foray into this new arena, we apply a few simple measures (derived from network science) to an observed galaxy distribution and explore whether these tools can be useful for cosmological and astrophysical studies. 

The structure of our paper is as follows. In \S2, we introduce some terminology and a simple recipe to construct a network from a given galaxy point distribution (derived from either a simulation or an observed catalog). In \S3, we apply the techniques to redshift slices of (a) the dark matter halo distribution from the Millennium-II simulation (Boylan-Konchin et al. 2008) and (b) the observed galaxy distribution from the Cosmological Evolution Survey (COSMOS; Scoville et al. 2013). We then present analyses of the latter. We discuss and summarize our results in \S4. 

The network computations are done using the free graph library $igraph$ (Csardi \& Nepusz 2006). 
Throughout, we adopt a $\Lambda$CDM cosmology defined by 
H$_0$ = 70 km s$^{-1}$ Mpc$^{-1}$, $\Omega_M = 0.3$, and $\Omega_\Lambda = 0.7$.

\section{Cosmic Networks}

In cosmological simulations, the full density distribution of matter is known; 
dark-matter halos (i.e., local density maxima) can be identified and 
 a cosmic network of matter distribution can be defined, that is closely related to the initial conditions and cosmological parameters. 
Topological features can be then identified using a variety of techniques. The smoothed density field can yield the Hessian matrix directly, from which wall and filamentary structures can be identified (e.g., Bond et al., Cautun et al. 2013). If only discrete halos are used, one can still employ a smoothing kernel or build a surface by triangulation (using, say, a cloud-in-cell scheme, e.g., Sahni et al. 1998, Sheth et al. 2003). 
 Likewise, in this paper,  we will identify topological structures using network measures.

In observational surveys, however, we are constrained by the nature of the observed galaxies. Galaxies are understood to be biased (and discrete) tracers of the underlying matter distribution. Observational surveys yield accurate positions in the plane of the sky; positions along the line of sight must be inferred using redshifts, which is subject to additional uncertainties. Here, we construct a network using the observed galaxy distribution, and therefore restrict ourselves to simple approaches that can be applied to observational data. The basic issue is the following: given a population of $n$ discrete galaxies, we need a simple algorithm to construct a network, and then a set of easily calculable measures that can robustly identify different topological structures.

In this section, we first introduce some basic network concepts in order to define terminology, describe a simple approach to constructing a cosmic network, and then demonstrate its application to two datasets, one theoretical and one observational. We assume that we are given a set of $n$ galaxies (or discrete halos) with known positions $\{\vec{x}_1, \vec{x}_2, \cdots, \vec{x}_n\}$.

\subsection{The Basics of Network Analysis}\label{sec:basic} 

Here, we briefly review the basic concepts used in network analysis before applying them to cosmic networks. 
We refer the interested reader to Newman (2003), Dorogovtsev and Goltsev (2008), and Barth\'elemy (2011) for further details. 

\subsubsection{Vertex, Edge, and Adjacency Matrix}

A \emph{network} or \emph{graph} is defined as a data structure composed of ``vertices'' connected by ``edges''. 
We denote the number of vertices by $n$ and the number of edges by $m$, following the notations in the mathematical literature. 

Edges have three properties which define the categories of networks: multiplicity, direction and weight.  The multiplicity is the number of edges between a given pair of vertices. If the multiplicity is 0 or 1 (i.e., simple connections only) and edges are only between two distinct vertices (i.e., self-loops, where a vertex connects to itself, are not allowed), the network is \emph{simple}.  Direction is used to analyze graphs where the connectivity direction is relevant, i.e., graphs where the connection from vertex $i$ to $j$ does not guarantee its reverse connectivity.  Our cosmic network is a {\it simple and undirected network}. Finally, scalar weights can be assigned to edges if necessary.  In this paper, we use both unweighted and weighted edges; in the latter case we consider only the simplest case where the edge weight is related to the distance between two vertices (i.e., the edge length). 

% When the number of edges for a pair is larger than 1, we call those edges \emph{multiedges}. 
% We can also allow self-loop connections for each single vertex, which is called \emph{self-edge} or \emph{self-loop}. 
% If there are neither multiedges nor self-loops in a network, 
% we call it a \emph{simple network}; otherwise, \emph{multigraph}. 

To represent the edge connections mathematically, 
we use the adjacency matrix, $A_{ij}$, defined for simple, undirected, and unweighted networks as : 
\begin{equation}\label{eq:def}
A_{ij} = \left\{ \begin{array}{ll}
	1 & \textrm{if there is an edge between i and j vertices,} \\
	0 & \textrm{otherwise.} 
	\end{array} \right.
\end{equation}
where $A_{ij}$ is an $n\times n$ matrix. Each i-th vertex can be represented 
by an n-dimensional unit vector of $\boldsymbol{e}_{i} \equiv \delta^{i}_{k}$, where $\delta^{i}_{j}$ is a conventional Kronecker delta. For simple and undirected networks:
(1) $A_{ij} = A_{ji}$, (2) $A_{ii} = 0$, 
and (3) $\sum\limits_{i,j} A_{ij} = 2 m$. The first symmetric relation holds for all undirected networks.  
The second relation of zero diagonal terms is due to no self-edge in a simple network. 
The third relation is a trivial normalization condition of the total number of edges. 
% 
% Without further assumptions, we can extend this definition of the adjacency matrix to directed multigraphs by assigning the number of 
% multiedges and self-edges to $A_{ij}$, while as a tradeoff we lose the two properties, $A_{ij} \neq A_{ji}$ and $A_{ii} \neq 0$.  
% In this paper, however, we restrict the use of the adjacency matrix  only for simple and undirected networks as defined in Equation~\ref{eq:def}. 

\subsubsection{Network Quantities Derived From the Adjacency Matrix}\label{sec:matrix}

Networks are represented by $n$,$m$ and $A_{ij}$.
Though network analysis itself heavily depends on numerical calculations, 
many network measures can be defined quite simply as analytical functions of $n, m,$ and $A_{ij}$.  
For example, degree centrality, $k_i$, defined as the number 
of connected neighbors for a given vertex $i$, can be written as 
\begin{equation}
k_i = \sum\limits_{j} A_{ij}, 
\end{equation}
because the i-th row (or column due to symmetry) of $A_{ij}$ represents the neighboring vertices to the i-th vertex.

If we denote the i-th vertex as a unit vector ${\delta}^{i}_{j}$, where ${\delta}^{i}_{j}$ is the conventional Kronecker delta 
and $j$ is the vector index, $j=1,\cdots, i,\cdots, n$, 
%and the adjacency matrix as $\mathbf{A}$, 
then the neighboring vertices to the i-th vertex, ${n}^{i}_{j}$, can be written as the multiplication
of  the unit vector by the adjacency matrix, 
\begin{equation}\label{eq:basic}
n^i_j = A_{jk} \delta^i_k, 
\end{equation}
where $k$ is a summation index. 
Recursive multiplications of Equation~\ref{eq:basic}
\begin{equation}
^{r}n_j^{i} =  A_{j k_r} \cdots A_{k_3 k_2}  A_{k_2 k_1}  \delta^i_{k_1} , 
\end{equation}
(where $r$ is the number of recursive multiplications)  
define paths in the network; since each multiplication of $A_{ij}$ moves the input vertices 
to their neighbors,  each element of the vector $^{r}n^{i}_j$ represents the number of possible paths 
connecting the j-th vertex from the i-th vertex by $r$ steps. 
We call such number of steps (edges) connecting two vertices as \emph{path length}. 
Then, the number of paths from the i-th vertex to the j-th vertex with the path length $r$, $N_{ij}^{r}$, 
is just the $ij$-th component of the recursively multiplied adjacency matrix, $\mathbf{A}^{r}$ : 
\begin{equation}\label{eq:ad}
N_{ij}^{r} =  [\mathbf{A}^{r}]_{ij}.
\end{equation}
The number of triangles in a network can be easily written using Equation~\ref{eq:ad} as : 
\begin{equation}\label{eq:tri}
\textrm{the number of } \bigtriangleup = \frac{1}{6} \sum_{i = j} N_{ij}^{3} = \frac{1}{6}  tr(\mathbf{A}^{3}),
\end{equation}
where the factor $\frac{1}{6}$ is due to 6 redundant path counts for a single triangle. 

\subsection{Building Networks}

Here, we present how to build a network from a generic population of $n$ objects with a spatial distribution, $\{ \vec{x}_1, \vec{x}_2, \cdots, \vec{x}_n \}$.  

%Therefore, it depends on the building recipe how effective the network analysis is when investigating cosmic structures. 
%In this paper, instead of choosing an exotic recipe, we choose a simple top-hat window linking all the neighbors 
%within the window. Then, we measure network quantities from the built network. 
%The following sections describe how to define the appropriate linking length and its relation with density contrasts. 

\subsubsection{Adjacency Matrix : Population and Linking length}

The simplest way to construct a network is to only link pairs that satisfy a distance criterion, where an edge is defined only if the edge length (i.e., the distance between two vertices) is less than a certain \emph{linking length} $l$; i.e.: 
% 
% Its corresponding adjacency matrix can be written as 
\begin{equation}\label{eq:lldef}
A_{ij} = \left\{ \begin{array}{ll}
	1 & \textrm{   if   } r_{ij} \le l,  \\
	0 & \textrm{   otherwise, } 
	\end{array} \right.
\end{equation}
where $r_{ij}$ is the distance between the two vertices, $i$ and $j$. 
Hence, when a ``population'' and a ``linking length'' are given, we can derive one unique network from them. 
One of the advantages using this definition is that the number of neighbors, degree centrality (DC; sometimes, simply called ``degree''),  
is just the source count within a volume of diameter 2$l$.  
Hence, the basic network statistic of DC is simply related to a local environmental density.
 Details regarding the degree, closeness and betweenness centralities are presented in \S \ref{sec:centrality}.

%The choice of linking length is arbitrary; however, since it defines the network, it is important to pick a value that has some physical meaning or statistical utility. For example, cosmological networks could be constructed using $8h^{-1}$Mpc 
%(since this is a typical scale on which the cosmic density fluctuation is close to unity), or some multiple of the typical virial radius of a given population, or a value that adequately samples the range of density based on the population size available.

\begin{figure*}[t]
\centering
\includegraphics[height=6.5 in]{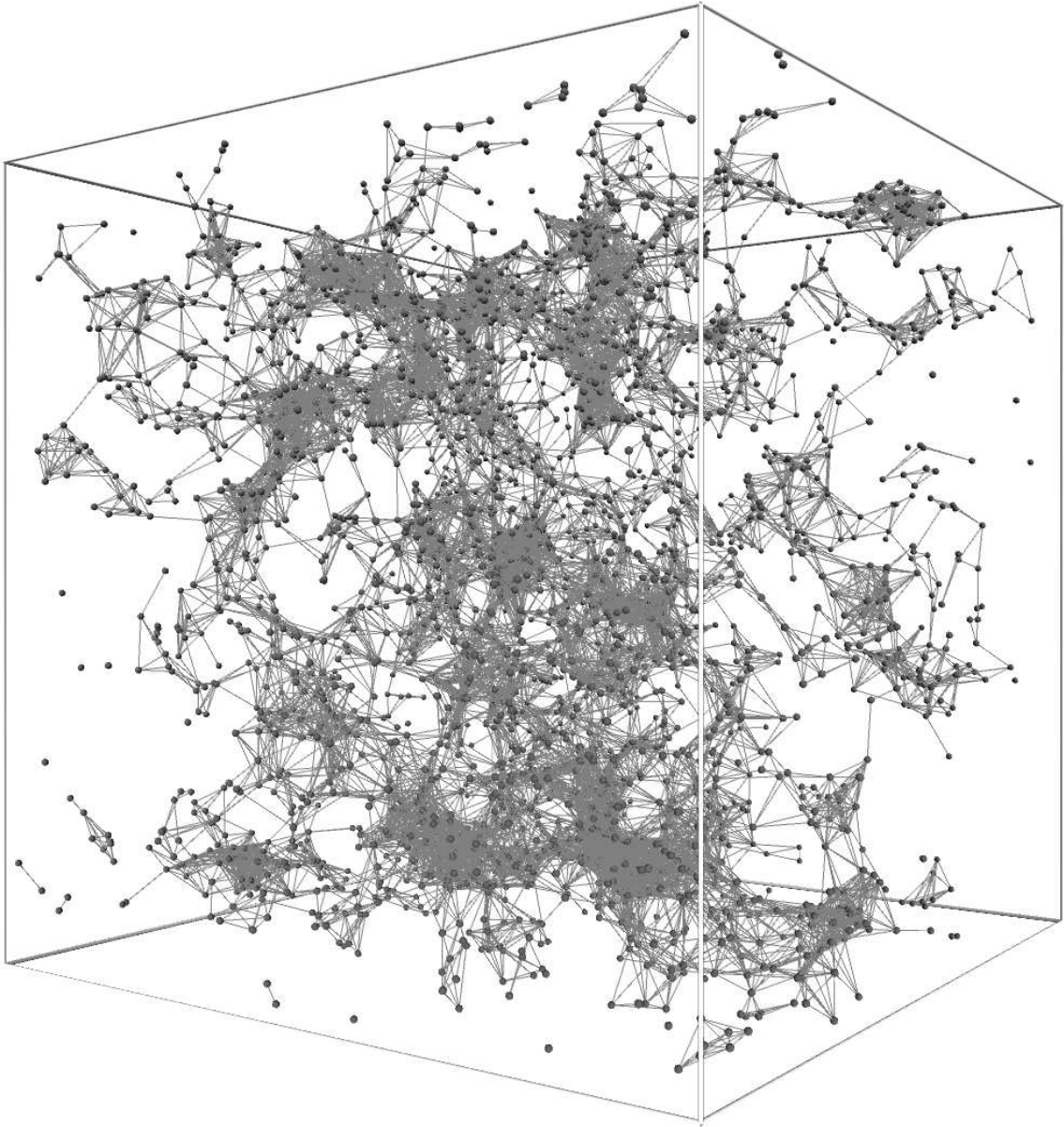}
\caption{The dark matter halo network made by the top 3375 ($= 15^3$) massive 
halos using the linking length $l_6$ ( $= 7.6 h^{-1}$ Mpc) at  z=3.06 from Millennium-II Simulation (Boylan-Konchin et al. 2008). 
The box is $100 h^{-1}$ comoving Mpc on a side. This three-dimensional visualization was constructed using 
the S2PLOT progamming library (Barnes \& Fluke 2006). 
}\label{fig:ms2show}
\end{figure*}

\begin{figure*}[t]
\centering
\includegraphics[height=6.0 in]{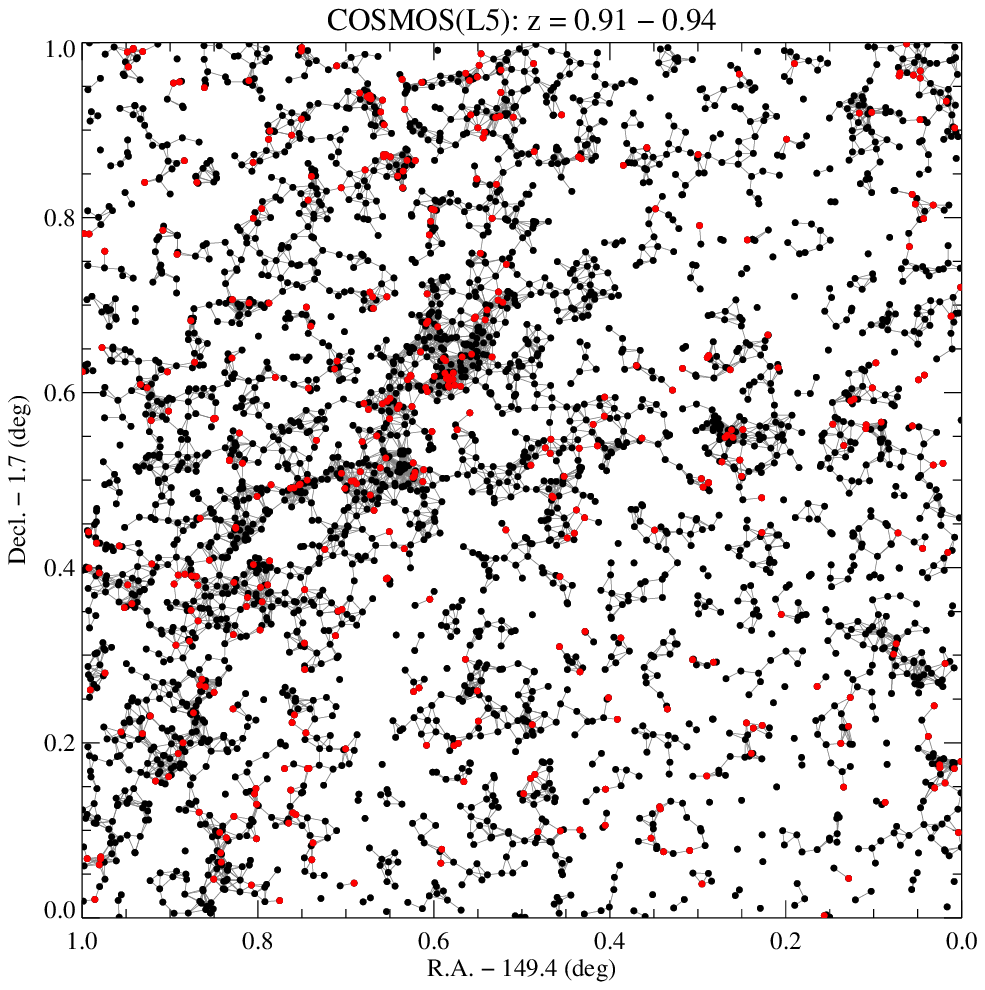}
\caption{ The two-dimensional network constructed using 3,366 galaxies in the photometric redshift slice $0.91<z<0.94$ 
from the COSMOS survey (Scoville et al. 2013). 
 The colors represent the ``quiescent'' (red) and ``star-forming'' (black) galaxies, 
selected by the two-color criteria (Ilbert et al. 2013), described in \S~\ref{sec:topology}. 
The linking length used is $0.0216^{\circ}$, which corresponds to 1.2 Mpc (adopting the cosmological parameters used in Scoville et al.). The field of view is $\approx1^\circ$ on a side, which corresponds to a comoving size scale of 54~Mpc. 
% The region size is $ \approx 1.0 ~ deg^2$ ranging R.A. $ = 149.4^{\circ} - 150.4^{\circ}$ and Decl. $ = 1.7^{\circ} - 2.7^{\circ}$.
\label{fig:cosmosshow}}
\end{figure*}

\subsubsection{Random Networks and Poisson Degree Distributions}

% Still, the choice of linking length is arbitrary. To set some criteria to choose linking length, 
% we investigate the relations between linking lengths and DC statistics for random networks as follows, 
% since random populations and their corresponding random networks are easy to calculate 
% and they are good comparison sets to real cosmic networks. 

 Before we construct a network from a real galaxy survey, we first investigate a random network of galaxies in order to illustrate the relationship between the choice of linking length and the resulting degree distribution (i.e., the measure of the local environmental density). A real galaxy survey is characterized by a sample of $n$ galaxies tracing an underlying cosmic matter density field $\rho(x)$ with some cosmic variance. In contrast, our random network is defined by $n$ trial samples (of ``galaxies'') drawn from an underlying probability density field $p(x)$ with a given Poisson variance. Random networks are well understood and relatively easy to generate and investigate, and their comparison with real galaxy distributions can be instructive. 

We assume that a point distribution is defined by a probability density function, $p(\vec{x})$, in a given survey volume, $V$, such that
\begin{eqnarray}
p(\vec{x})  &\equiv& \bar{p}~ ( 1 + \delta(\vec{x})) \label{eq:contrast}, 
\end{eqnarray}
where $\bar{p}$ represents the average probability and $\delta(\vec{x})$ the probability contrast. If we normalize the probability to unity, then 
\begin{eqnarray}\label{eq:norm}
\int_{V} p(\vec{x})~ dV &\equiv& 1, \\
\bar{p}  &=& \frac{1}{V} \label{eq:a},\\
\int_{V} \delta(\vec{x})~ dV &=& 0 \label{eq:b}. 
\end{eqnarray}
$V$ is not restricted to 3 dimensions; in a two-dimensional survey, $V$ represents the area. 
%
% We rewrite the probability density function with its average probability, $\bar{p}$, and its probability contrast, $\delta(\vec{x})$,
% \begin{eqnarray}
% p(\vec{x})  &\equiv& \bar{p} ( 1 + \delta(\vec{x})) \label{eq:contrast}, \\
% \bar{p}  &=& \frac{1}{V} \label{eq:a},\\
% \end{eqnarray}
% where Equation ~\ref{eq:a} and ~\ref{eq:b} are derived from Equation ~\ref{eq:norm}. 
%
% In these settings, the random population does not have to be a uniform flat random; i.e., $\delta(\vec{x}) \neq 0$.  
% the volume represents two-dimensional volume, i.e., area. 

For a top-hat volume, $V^i_l$, centered at $\vec{x}_i$, 
the mean probability density, $p_i$, and its mean probability contrast, $\delta_i$, are defined as:
\begin{eqnarray}
p_i  &\equiv& V_l^{-1} \int_{V^i_l}  p(\vec{x})~ dV, \label{eq:mean}  \\
\delta_i  &\equiv& V_l^{-1}  \int_{V^i_l}  \delta(\vec{x})~ dV,  
\end{eqnarray} 
 where $V_l$ is the size of top-hat volume, $V^i_l$.
From Equation \ref{eq:contrast}, we can rewrite Equation \ref{eq:mean} as  
\begin{eqnarray}
p_i &=& V_l^{-1} \int_{V_l} \bar{p}  ( 1 + \delta(\vec{x})) dV, \nonumber \\
 &=&  \bar{p} (1 + \delta_i).
\end{eqnarray}
Since $p(\vec{x})$ is a probability density, the real probability, $\mathbb{P}_i$, falling in the volume, $V_l$, centered at $\vec{x}_i$ is 
\begin{eqnarray}
\mathbb{P}_i &\equiv&  p_i V_l. 
\end{eqnarray}
For a random ensemble, each realization is a binomial trial. For $n-1$ trials, therefore, the number of data points 
falling in the top-hat volume is 
\begin{eqnarray}
P_i (k) &=& \frac{(n-1)!}{k! (n-1-k)!}~ \mathbb{P}_i^k (1-\mathbb{P}_i)^{n-1-k} , \label{eq:bi} \\
\mu_i &=& (n-1) \mathbb{P}_i, \label{eq:bi2}
\end{eqnarray}
where $\mu_i$ is the mean counts. We derive Equation \ref{eq:bi} and \ref{eq:bi2} for $n - 1$ trials at a position, $\vec{x}_i$, instead of $n$ 
trials at a random position. The case of $n - 1$ trials is the statistic of neighbor counts for each vertex, 
more relevant to the network analysis discussion in the next section, while the case of $n$ trials is the source count statistic at an arbitrary window position. 
For a large $n$, there is little difference between the two, and Equation ~\ref{eq:bi} asymptotes to a Poisson distribution 
\begin{eqnarray}
P_i(k) &\sim& \frac{{\mu_i}^k e^{-\mu_i}}{k!} \label{eq:poisson}.  
\end{eqnarray}

 Finally, we connect this Poission distribution with the probability contrast as follows.
If $p(\vec{x})$ is uniform (i.e., $\delta(\vec{x}) \equiv 0$), its probability,$\bar{\mathbb{P}}$, and mean counts, $\bar{\mu}$, can be written as  
 \begin{eqnarray}
\bar{\mathbb{P}} & = &  \bar{p} V_l \nonumber \\
& = & \frac{V_l}{V},\label{eq:pbar} \\
\bar{\mu} &=& (n-1)\frac{V_l}{V} \label{eq:mubar}.
\end{eqnarray}
 From these, we define the probability contrast at each position, $\delta_i$, in terms of $\mu_i$ and $\bar{\mu}$,  
\begin{eqnarray}
\mu_i &\equiv& \bar{\mu}~ ( 1 + \delta_i ), \\
\delta_i &=& \frac{\mu_i - \bar{\mu}}{\bar{\mu}} \label{eq:overdense}.   
\end{eqnarray}
Equation~\ref{eq:poisson}--\ref{eq:overdense} are derived by assuming 
that the given population is a random realization of an underlying probability function, $p(\vec{x})$. 
 Hence, random networks are well characterized by Poisson distributions and their mean values.

When $p(\vec{x}) \equiv \bar{p}$ (i.e., $\delta_i \equiv 0$), 
the degree (DC) distribution follows the Poisson shot noise statistics of Equation~\ref{eq:poisson} 
with the mean value of Equation~\ref{eq:mubar} (Erd\H{o}s and R\'{e}yni 1959).  
For a general $p(\vec{x})$ (i.e., $\delta_i \ne 0$), 
%the DC distribution is a convolution of {\bf the $\delta_i$  distribution} (Equation~\ref{eq:overdense}) with its corresponding Poissonian distribution (Equation~\ref{eq:poisson}). 
 the DC distribution is a sum of all Poisson distributions (Equation~\ref{eq:poisson}) for their corresponding $\delta_i$ (Equation~\ref{eq:overdense}).
For $\delta = -0.8$ and $3.0$, the mean counts are $0.2\bar{\mu}$ and $4.0\bar{\mu}$, 
and their Poisson distributions are $P_{\mu_i = 0.2 \bar{\mu}}(k)$ and $P_{\mu_i = 4.0 \bar{\mu}}(k)$. 
If we choose a linking length, $l$, to make $\bar{\mu} = 5$, the regions where $\mu = 1$ (or $\mu = 20$) 
correspond to density contrasts of $\delta = -0.8$ (or $\delta = 3.0$). 
Therefore, for random networks, the linking length defines all of Equation~\ref{eq:poisson}--\ref{eq:overdense}. 

While random networks are well known to exhibit Poisson degree distributions (Erd\H{o}s and R\'{e}yni 1959), the degree distribution of linked web sites in the World-Wide Web exhibit a power-law, or scale-free, distribution (Albert et al.~1999). This was a surprising result, especially given the autonomous growth of the WWW without any central authority controlling the creation and linking of web documents. Scale-free networks appear to be a common phenomenon in nature, and have now been seen in the networks of scientific papers' citations, co-starring of movie actors, and protein-protein interactions (Barab\'asi 2009). 

\subsubsection{Linking Length}
To derive the specific relations among $n$, $l$, and $\bar{\mu}$, we write the spherical linking volume as  
\begin{eqnarray}
V_l = \alpha_N l^N, 
\end{eqnarray}
where $N$ is the number of dimensions and $\alpha_N$ is the volume for a unit radius; i.e.,  $\alpha_2 = \pi$ and $\alpha_3 = \frac{4}{3}\pi$.
From Equation~\ref{eq:mubar}, the linking length, $l$, can be reduced to    
\begin{eqnarray}
l   &=& \bigg[ \frac{\bar{\mu} V}{ \alpha_N (n-1)} \bigg]^{1/N} \label{eq:linkinglength}.  
\end{eqnarray}
Specifically, if the survey dimensions are N=2 or 3 with square or cubic survey volumes, the linking lengths are  
\begin{equation}
l = \left\{ \begin{array}{ll}
	\bigg[ \frac{\bar{\mu}}{ \pi (n-1)} \bigg]^{\frac{1}{2}} L & \textrm{for } N = 2 \textrm{  and } V = L^2 \\
	\bigg[ \frac{3\bar{\mu}}{ 4\pi (n-1)} \bigg]^{\frac{1}{3}}L & \textrm{for } N = 3 \textrm{  and } V = L^3
	\end{array} \right., \label{eq:ll}
\end{equation}
where $L$ is the system size of the survey region.

In general, the sample size, $n$, and the survey volume, $V$, are known.  
$\bar{\mu}$ and $l$ can be determined by fixing one of them. 
For example, if we take $\bar{\mu} = 5$, its corresponding linking length can be determined by Equation~\ref{eq:linkinglength}.   
The Poisson distributions, $P_{\bar{\mu}=1}(k)$, $P_{\bar{\mu}=5}(k)$, and $P_{\bar{\mu}=20}(k)$,
represent the random variation of the neighbor counts, $k$, for $\delta = -0.8, 0.0,$ and $3.0$. 
Alternatively, if we choose $l = 8 h^{-1} $ Mpc from a cubic survey volume, $[100 h^{-1}$ Mpc$]^3$, in the comoving scale,  
its corresponding $\bar{\mu}$ can be calculated from Equation~\ref{eq:linkinglength}. 
The Poisson distributions for $0.2\bar{\mu}$, $\bar{\mu}$, and $4.0\bar{\mu}$
represent the random variances of the neighbor counts, $k$, for $\delta = -0.8, 0.0,$ and $3.0$.
We denote this dependence by $l_{\bar{\mu}}$. When $\bar{\mu} = 5$, we denote the linking length by 
$l_5$  and its corresponding Poisson distribution by $P_5(k)$. Then, $P_1(k)$ and $P_{20}(k)$ are 
the Poisson distributions for $\delta = -0.8$ and 3.0 accordingly. 

We use a fixed linking length to construct networks for two cases. 
Figure~\ref{fig:ms2show} shows the network resulting from the distribution of the most massive 3375 ($= 15^3$) dark matter halos at a redshift $z=3.06$ from the Millennium-II simulation, constructed  
using the linking length $l_6$ ( $= 7.6h^{-1}$ Mpc). 
The length of cubic box is 100$h^{-1}$ comoving Mpc. 
Figure~\ref{fig:cosmosshow} shows the network resulting from the distribution of 3366 galaxies in the photometric 
redshift range $0.91 < z < 0.94$ 
from the COSMOS survey data (Scoville et al., Ilbert et al. 2013). Galaxies have been divided by their optical colors:  
``quiescent'' (red) and ``star-forming'' (black) galaxies, selected by the two-color criteria from Ilbert et al., described in \S~\ref{sec:topology}. 
The linking length is $0.0216^{\circ}$, corresponding to 1.2 comoving Mpc, when adopting the cosmological parameters used in Scoville et al. 
The region size is $ \approx 1^\circ \times 1^\circ$, corresponding to a comoving size scale of $54 \times 54~ \textrm{Mpc}^2$. The survey covers the range R.A. $ = 149.4^{\circ} - 150.4^{\circ}$ and Decl. $ = 1.7^{\circ} - 2.7^{\circ}$.

  The linking length is a free parameter in our study, analogous to the choice of the size scale of smoothing kernels in environmental studies using traditional density measures. The choice of the ideal linking length will depend on the number of points in the network and the desire to create meaningful connections without connecting too few or too many galaxies in the network. If the linking length is chosen to be too small, most objects are isolated; similarly, if the linking length is too large, all galaxies can be connected to form a complete graph. In both extremes, the derived network quantities will not permit separating the galaxies into useful topological classes in which we can compare their properties. The linking length can also be chosen using physical intuition. For a galaxy network, the linking length should sample intergalactic scales that probe the observed large scale structures and can, for example, separate galaxy clusters from filamentary regions.

For the COSMOS data, we find that the linking length $l_5$ (corresponding to 1.2 comoving Mpc) is a practical and physically acceptable scale. In this pilot study, we investigate this COSMOS-$l_5$ network and present the results obtained from network analysis. We investigated $l_4$, $l_6$, and $l_7$ networks and found that $l_4$ results in a relatively isolated network with less filamentary structure, whereas $l_7$ begins to over-connect the network; $l_5$ and $l_6$ are qualitatively similar. In \S \ref{sec:future}, we discuss possible caveats of this recipe to build networks using linking length and suggest future improvements.

\section{Results}
We have presented the general ideas of network analysis and 
defined the linking recipe and its related random Poisson distribution to build a network structure from given $n$ data points.  
In this section, we apply the network analysis tools to the cosmic network derived for the $0.91<z<0.94$ galaxies (Figure~\ref{fig:cosmosshow}) and discuss the properties of galaxies in the resulting topological classes. 
Since Millenniuum-II Simulations can provide the halo properties only, 
we focus on the observed COSMOS galaxies and explore whether topology has any effect on galaxy 
color, star-formation rate (SFR) and stellar mass. 

\subsection{Topological Selections : Centrality}\label{sec:centrality}

% Since a person having most friends in a social community can be considered as the most important person, 

A major advantage of using network analyses is that we can utilize various topological measures, called ``centrality'', assigned to each vertex indicating the vertex's importance for a given topological feature.  For example, in social networks, degree centrality, the number of neighbors for each vertex, represents the number of friends; i.e.,  the DC measures social importance.\footnote{The degree of a network is a very simple measure, and more sophisticated measures may result in better results. For example.  Page et al. (1999) suggest a variant of DC, ``PageRank'' (PR), to prioritize the importance of web pages searched using Google.  While DC gives an equal weight ``1'' to each neighbor (hence, the summation of neighbors' weights is equal to the number of neighbors), the PR method assigns a different weight on each vertex (See Page et al. for details).  This modification resulted in better ranks for the importance of WWW, suggesting that the the intention of people's search queries are topologically more related to the PR measurement rather than the DC measurement.}

Here, we focus on three simple centrality measures to analyze the network: 
Degree Centrality (DC; \S\ref{sec:dc}); 
Betweenness Centrality (BC; \S\ref{sec:bc}); 
and Closeness Centrality (CL; \S\ref{sec:cl}). 
These have the benefits of mathematical simplicity and ease of interpretation, and thus serve as a useful first step in our 
enterprise of exploring the utility of network analyses tools in astrophysics. 

\begin{figure*}[t]
\centering
\includegraphics[height=6.0 in]{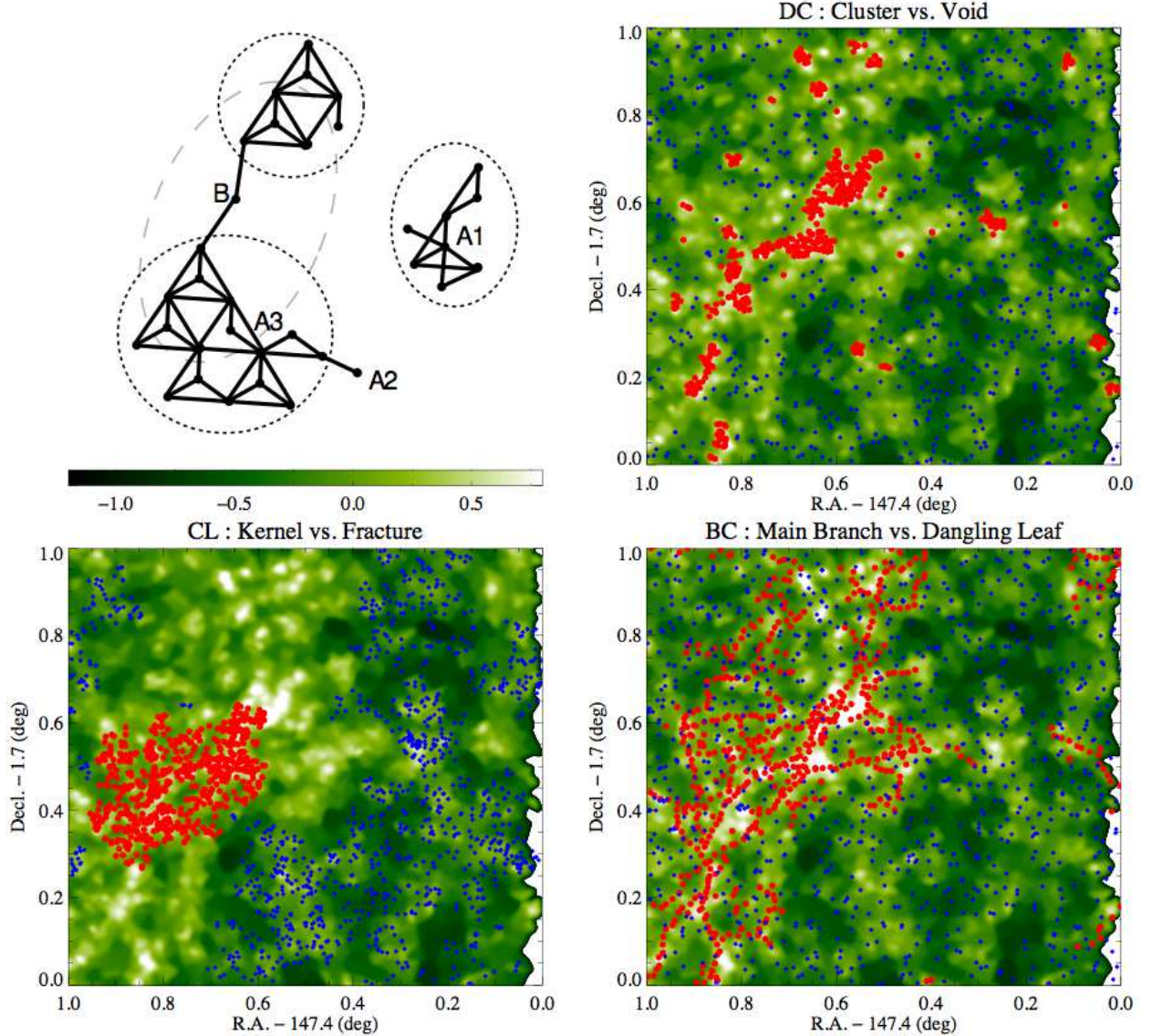}
\caption{ This figure illustrates the galaxy selections resulting from the network-based measures discussed in the text. 
The schema in the top left panel illustrates the topological meaning of the three main network-based selections, DC, BC and CL. The points represent vertices (galaxies) and the solid lines are edges linking the vertices. The vertices A1 and A3 are both well connected to many neighbors and have high DC values (i.e., they lie in a ``Cluster"). B is a vertex with high BC (i.e., a ``Filament" or ``Main Branch" galaxy) since the main path between highly connected regions pass through it.  A2 is a vertex with BC=0 lying on the edge of a dense cluster, and represents a ``Dangling Leaf" galaxy. While A1 and A3 are both in Clusters, the region of A1 is unconnected to the main backbone structure, and it therefore belongs to a ``Fracture" with low CL. The remaining 3 panels illustrate the DC, CL and BC selections using the galaxies lying at redshifts $0.91<z<0.94$ in the COSMOS field (Scoville et al. 2013). 
The green scale shows the logarithmic density distribution of galaxies (in units of ${\rm log(\rho/\bar{\rho})}$), where $\bar{\rho}$ is the average value of the density field. 
}\label{fig:compdensity}
\end{figure*}

The three measures are illustrated in the top left panel of Figure~\ref{fig:compdensity}. In the following subsections, we discuss each of these measures in turn, define topological classes of galaxies based on the ranges of these measures, and analyze whether galaxy properties differ between these classes. We also investigate whether some measures are better than others in predicting galaxy properties, and if so, what this implies for topological studies of galaxy evolution. 

% %In our linking recipe for galaxies (or dark matter halos) in a cosmic survey 
% %(or a cosmological simulation), 
% Since the DC measurement is consistent with the conventional source count overdensity, 
% in the aspect of network analysis, 
% we have only used the DC measurement when investigating cosmic webs. 
% As the PR measurement works better for the WWW structure, 
% there could be better centralities (i.e., better topological features) to understand 
% the structure of cosmic web.
% In this paper, we present three measures: 
% %, and investigate the relations between their measures and galactic properties. 
% The top-left panel of Figure~\ref{fig:compdensity} demonstrates the topological meanings of DC, BC, and CL. 
% We apply each to the cosmological distributions discussed above 
% and assess the utility of these measured in analyzing cosmic structures.
% The studies using other centralities (or newly invented centralities) are left for the future.  

% BC is a measure of the importance of a vertex in a network, as defined by the number of connected paths that pass through it. 

\subsubsection{Void, Wall, and Cluster : Degree Centrality}\label{sec:dc}

\begin{figure*}[t]
\centering
\includegraphics[height=3.0 in]{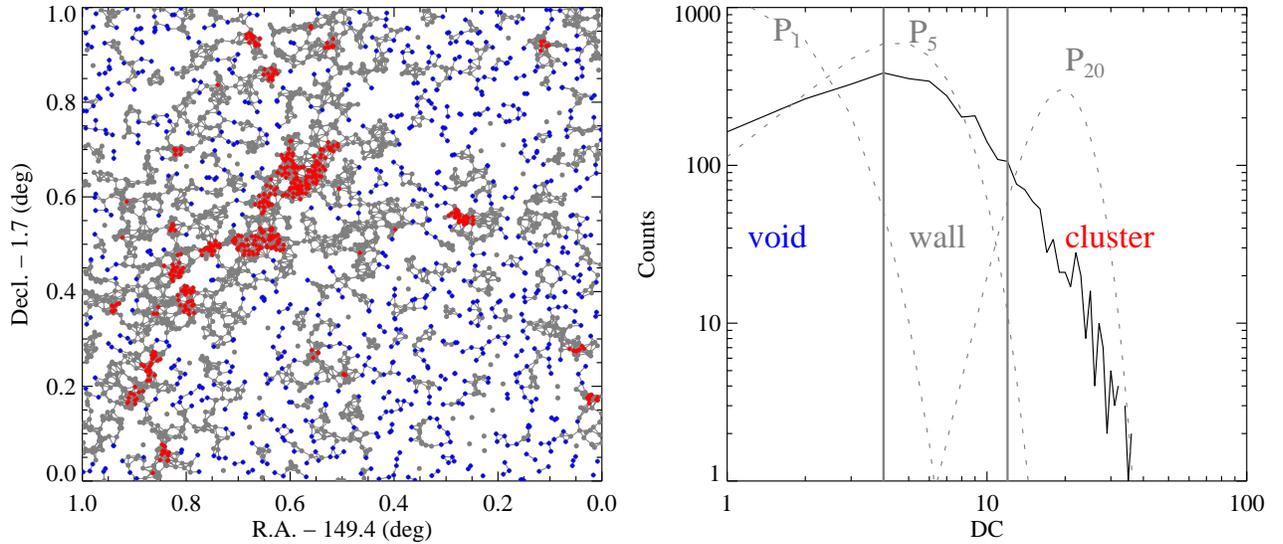}
\caption{The spatial distribution (left panel) and DC distribution (right panel) for the COSMOS network shown 
in Figure~\ref{fig:cosmosshow}. The different colors in the left panel represent galaxies that lie in Void (blue), Wall (grey) and Cluster (red) topological regions, as defined by the vertical lines in the right panel. 
This result is comparable to the galaxy density derived by Voronoi tessellation (shown in Figure 6--7 of Scoville et al. (2013) and in the color contours in Figure~3 of this paper)  
since both trace local population density.  
}\label{fig:dgcosmos}
\end{figure*}

\begin{figure*}[t]
\centering
\includegraphics[height=3.0 in]{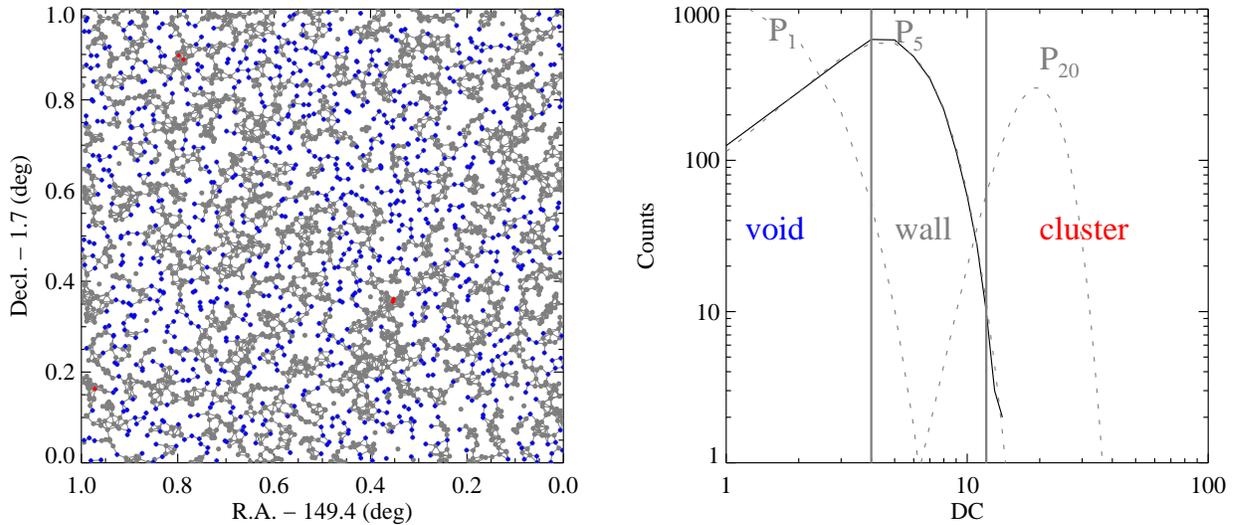}
\caption{Similar to Figure~\ref{fig:dgcosmos}, but for a random population. 
Since we choose the linking length $l_5$, the DC distribution for this random population exactly follows the $P_5$ distribution. 
}\label{fig:dgrandom}
\end{figure*}

DC, the degree of a vertex, is simply the number of linked neighbors; in Figure~\ref{fig:compdensity}, the degrees of the vertices A1 and A2 are 5 and 1 respectively.  In the cosmic network, DC is the number of galaxies within a top-hat window of radius $l$ (the linking length). Figure~\ref{fig:dgcosmos} shows both the spatial distribution of galaxies (left panel) and the DC distribution resulting from the COSMOS network (right panel). The right panel also shows the Poisson distributions that would result for random distributions corresponding to P$_5$ (i.e., $\delta=0.0$, or a random distribution with average counts; see Figure~\ref{fig:dgrandom} for an example), P$_1$ (an underdensity of $\delta=-0.8$), and P$_{20}$ (i.e., an overdensity of $\delta = 3.0$).

% The right panel of Figure~\ref{fig:dgcosmos} shows the DC distribution of the COSMOS network 
% shown in Figure~\ref{fig:cosmosshow}. Since the linking length is $l_5$, the average (i.e., $\delta=0.0$) random source counts 
% follow the Poisson distribution, P$_5$. The corresponding random distributions for 
% $\delta = -0.8$ and $\delta = 3.0$ are P$_1$ and P$_{20}$. Figure~\ref{fig:dgrandom} shows 
% the same figure with Figure~\ref{fig:dgcosmos} for a random population. 
% Due to the definition of $l_5$, the DC distribution of this random network follows exactly the P$_5$ distribution. 

Based on this $l_5$ network,  
we divide the vertices (galaxies) into three topological classes according to their DC values :
\begin{itemize}
\item Void : DC $< 4$,
\item Wall : $4 \le $ DC $ \le 12$, 
\item Cluster : $12 < $DC.    
\end{itemize}
We chose the thresholds of ``4'' and ``12'' based on the related Poisson probabilities,  
$\sum\limits_{k \ge 4} P_1(k) = 0.019$, $\sum\limits_{k \ge 3} P_1(k) = 0.08$, 
and $\sum\limits_{k \le 12} P_{20}(k) = 0.039$. Roughly, 95\% of the random ensembles  
for $\delta = -0.8$ and $\delta = 3.0$ fall in ``Void'' and ``Cluster'' respectively. 
This selection is not unique since the real cosmic ensemble is not random  
and the Poisson shot noise is still large within the $l_5$ window.  
 The choice of limits for the centrality measures will depend on the network that is constructed. 
The limits chosen in our paper are appropriate for the sample size and network defined 
by the COSMOS data. A much denser galaxy sample in the same volume will likely require different limits. 
Therefore, these three classes (defined by DC) represent
a qualitative selection based on the local population density.

%\clearpage
\subsubsection{Main Branch and Dangling Leaf : Betweenness Centrality} \label{sec:bc}

Betweenness Centrality is a measure of how many shortest paths (geodesic paths in graph) of all pairs 
pass through a certain vertex. This measure can be explained as traffic loads on a road network. 
If there is only one road connecting two large cities, all cars need to pass through this road to get to the other city. 
The vertices lying on this single pathway have higher BC values than other vertices. 
%The top-left panel of Figure~\ref{fig:compdensity} is a schematic figure demonstrating the topological meanings of DC, BC, and CL.
In the top-left panel of Figure~\ref{fig:compdensity}, 
the vertex B is an example of high BC but low DC: while it has only two neighbors, all the shortest pathways between the two clumps pass through it.
In the context of the BC measure,  the vertex B is more topologically important than other vertices with high degree. 
Therefore, BC is a promising topological measure to trace filamentary structures bridging large clusters. 

The Betweenness Centrality, $x_i$ for the $i$-th vertex, is defined as:
\begin{equation}
x_i = \sum\limits_{st} \frac{n^i_{st}}{g_{st}}, 
\end{equation}
where $g_{st}$ is the number of shortest paths between the vertices $s$ and $t$, 
and $n^i_{st}$ the number of these which pass through the vertex, $i$. 
If $g_{st}$ is zero, we assign $n^i_{st}/g_{st} = 0$. 
This BC definition can be applied to both weighted and unweighted networks, denoted as wBC and BC respectively. 
For our cosmic networks, we assign the real edge distance as a weight to each edge. 
We also define a weighted Degree Centrality (wDC) by giving additional weights to each vertex as 
\begin{equation}
\tilde{k}_i = k_i + \sum\limits_{j=1}^{k_i} \frac{L_{link} - l_{ij}}{L_{link}}, 
\end{equation}
where $L_{link}$ is the linking length, $k_i $ DC and $\tilde{k}_i $ weighted DC for the vertex, $i$,  
and $l_{ij}$ the edge distance between two vertices, $i$ and $j$.  
Among possible BC measurement variants, BC, wBC, and wBC/wDC, we find 
that wBC/wDC is the best measure among the three to trace filamentary structures. 

In the previous section we used DC measures to define Void, Wall and Cluster galaxy members. 
Here, in an analogous manner, we define two topological populations using BC measurements: 
``Main Branch" (or high BC) and ``Dangling Leaf'' (zero BC) galaxies. The Main Branch population 
traces the main connected structures of the galaxy distribution. 
The Dangling Leaf galaxies are unconnected to the denser regions and typically lie on the outer boundary  
of the galaxy distribution (as exemplified by the vertex A1 in the top-left panel of Figure~\ref{fig:compdensity}).

\begin{figure*}[t]
\centering
\includegraphics[height=3.0 in]{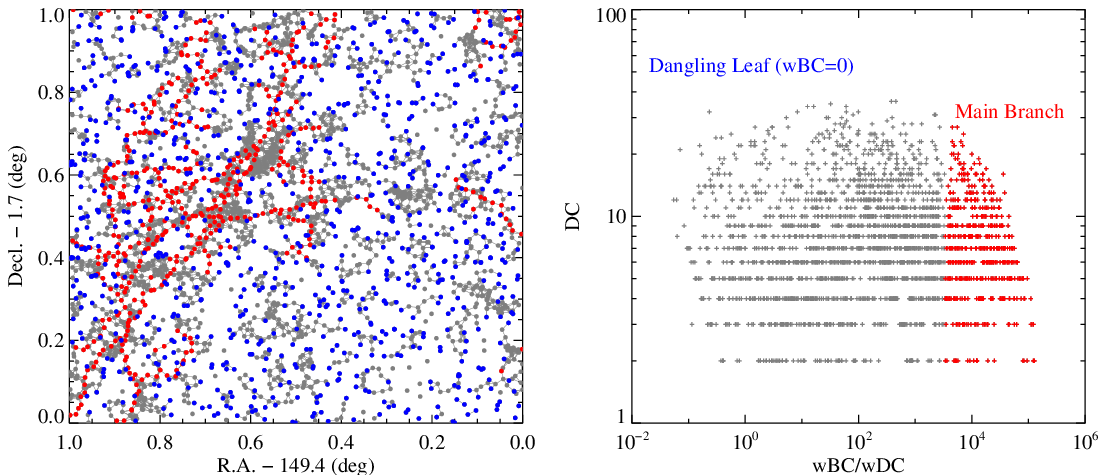}
\caption{The spatial distribution (left panel) showing Main Branch (red) and Dangling Leaf (blue) galaxies, and the distribution of wBC/wDC vs. DC (right panel) for the COSMOS galaxy sample. The Main Branch galaxies trace the filamentary structures well.
}\label{fig:bccosmos}
\end{figure*}

The DC selection described in the previous subsection resulted in 492 galaxies in the Cluster class. Hence, for comparison, we define the Main Branch to be the set of 500 galaxies with the highest BC values. There are 917 Dangling Leaf galaxies. 
% To compare this population with Cluster galaxies,  we choose the top 500 BC galaxies as Main Branch galaxies; the number of Cluster galaxies is 492. 
Figure~\ref{fig:bccosmos} shows the spatial positions of Main Branch (red) and Dangling Leaf (blue) and 
the distribution of wBC/wDC vs. DC for the COSMOS galaxy sample. 
The Main Branch galaxies trace the filamentary structures of the COSMOS network well. 

%\clearpage
\subsubsection{Fracture, Backbone, and Kernel : Closeness Centrality} \label{sec:cl}

CL is a measure of topological center, defined as the inverse of the average shortest distance from a given vertex to all the other vertices. Here, distance between pairs is measured by crawling on the network along edges, not using the straight paths connecting the pairs. 
Therefore, the vertex of the highest CL is connected to the other vertices by the shortest path-length on average. 
In other words,  
any influence or information at this highest CL vertex can spread most effectively to all the other vertices.
This topological center generally does not coincide with the highest DC.

% The mathematical definition of CL is written as 
CL is defined as:
\begin{equation}
C_i = \Big[ \frac{1}{n-1} \sum\limits_{j (\ne i)} d_{ij} \Big]^{-1},
\end{equation}
where $d_{ij}$ is the shortest path from $i$ to $j$ on the network. The term within the square bracket is the average shortest distance. 
Hence, the vertex with the highest CL has the smallest average distance. 
If there is no path to connect between $i$ and $j$ (i.e., unreachable pair), the pair's topological distance is infinite. 
Hence, we assign a sufficiently large value which is an upper bound for the pair distances of $d_{ij}$. 
In general, for unweighted networks, the number of vertices $n$ is used as the upper bound, 
since no shortest path can be larger than $n$. 

\begin{figure*}[t]
\centering
\includegraphics[height=6.0 in]{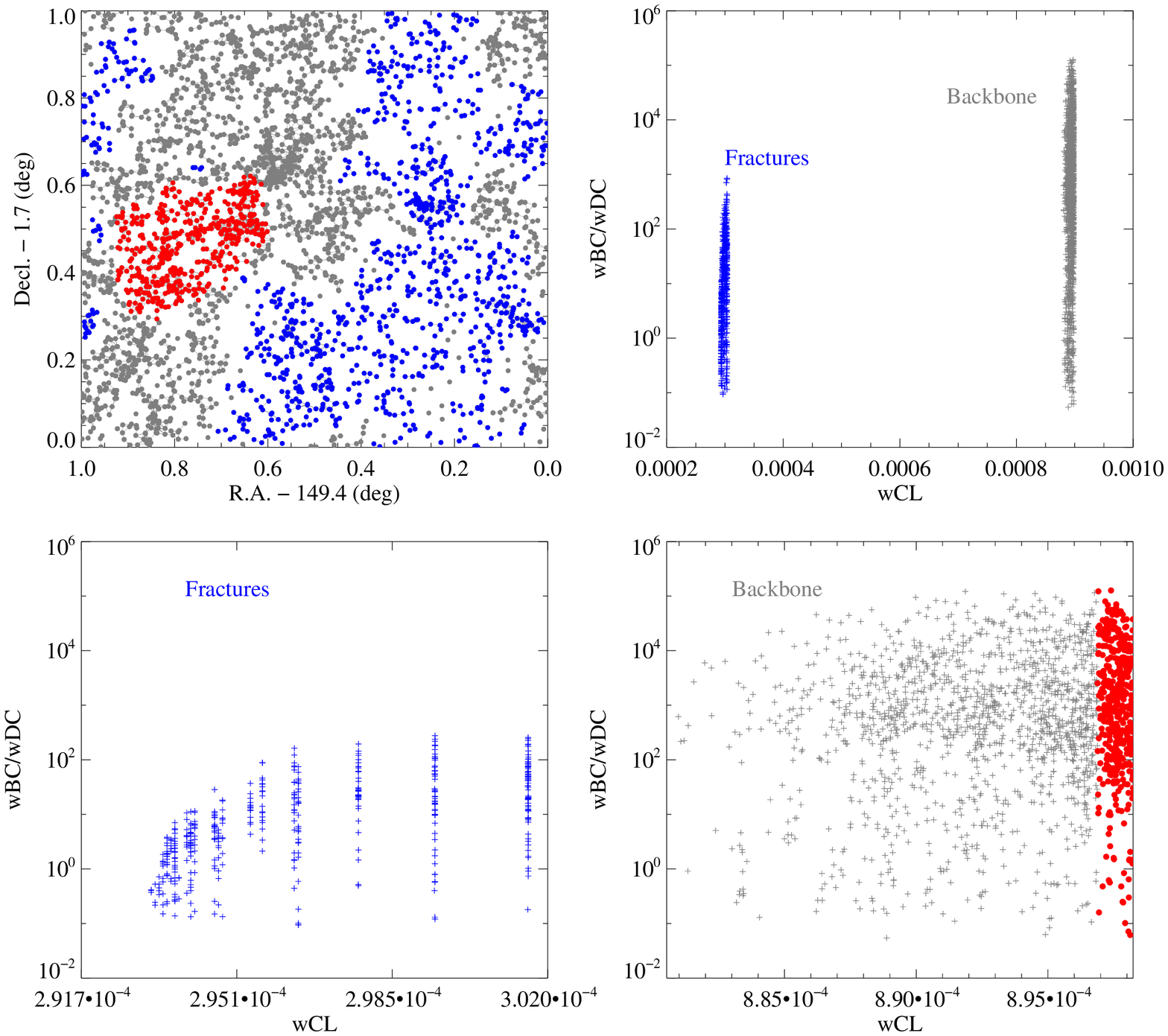}
\caption{The top-right panel shows the distribution of wBC/wDC vs. wCL 
and the top-left panel shows the spatial distributions 
of galaxies in the Fracture (blue), Backbone (grey), and Kernel (red) classes. 
The bottom panels show the zoom-in CL values for Fracture (bottom-left) 
and Backbone (bottom-right). When assuming the conventional walls, filaments, and clusters in the Universe 
are connected without discontinuity forming a single colossal backbone, 
Fracture can be used as a new topological definition of ``void''. 
}\label{fig:clcosmos}
\end{figure*}

Due to this artificial assignment of a large distance to unreachable pairs, 
the CL values show a bimodal distribution with values separated by large gaps (see top-right panel of Figure~\ref{fig:clcosmos}). 
We call the largest structure the ``Backbone''\footnote{This is generally referred as a ``giant component'' in the network terminology.} 
of the distribution, and refer to the other sub-clumps as ``Fractures''. 
If we assume that all walls, filaments, and clusters in the Universe are connected forming 
a single colossal Backbone, Fractures are analogous to void regions.  
Like the Main Branch in BC measurement, we choose the top 500 CL galaxies and call them ``Kernel''. 
Fracture, Backbone, and Kernel are comparable to Void, Wall, and Cluster, with different topological meanings. 
Specifically, by definition, Wall and Cluster are exclusive selections, having no intersection between them, 
while Kernel is a subset of Backbone. Hence, we also define an ad-hoc selection ``BackboneSub", excluding Kernel galaxies from Backbone. 
DC measures represent ``local environment'', while CL measurements 
represent ``topological and global environment''. 

The bottom panels of Figure~\ref{fig:clcosmos} show the zoom-in CL values for Fracture (bottom-left) 
and Backbone (bottom-right). We can observe more sub-fractures separated in CL values within Fracture. 
The top-left panel shows the spatial distribution for Fracture (blue), Backbone (grey), and Kernel (red). 
The spatial distributions of Fracture, Backbone, and Kernel are very different from Void, Wall, and Cluster, 
reflecting their different topological selections. 
%Due to the CL's global-scale selection, Kernel is a selection of (almost) all galaxies within $\approx 0.2$ degree diameter  
%and Fracture over 0.4 degree. 
%Generally these scale are large to smear out most physical properties to cosmic averages. 

 Unlike the local density measures, the BC and CL measures, which depend on the shortest path ways,  can be qualitatively affected by random noise. In particular, noise can result in bridging fractures which may, in reality, be separate regions.  These ``random bridges'' do not, however, strongly affect CL, since the latter is an average quantity for all possible pairs; only  low CL vertices in small fractures are sensitive to this random noise.  For BC, a random bridge can make a Dangling Leaf jump up to a Main Branch. However, the statistics of Dangling Leaves are not significantly affected by random bridges, 
since the surface area of the matter distribution in regions where Dangling Leaves reside is much larger than the possible junction spots 
of random bridges. Also, the regions that are newly added to the Main Branch as a result of the random bridges tend to be at the termini, and hence do not dominate the existing Main Branch.

%\clearpage
\subsection{Topological Classes and their Galactic Properties}\label{sec:topology}

\begin{figure*}[t]
\centering
\includegraphics[height=4.8 in]{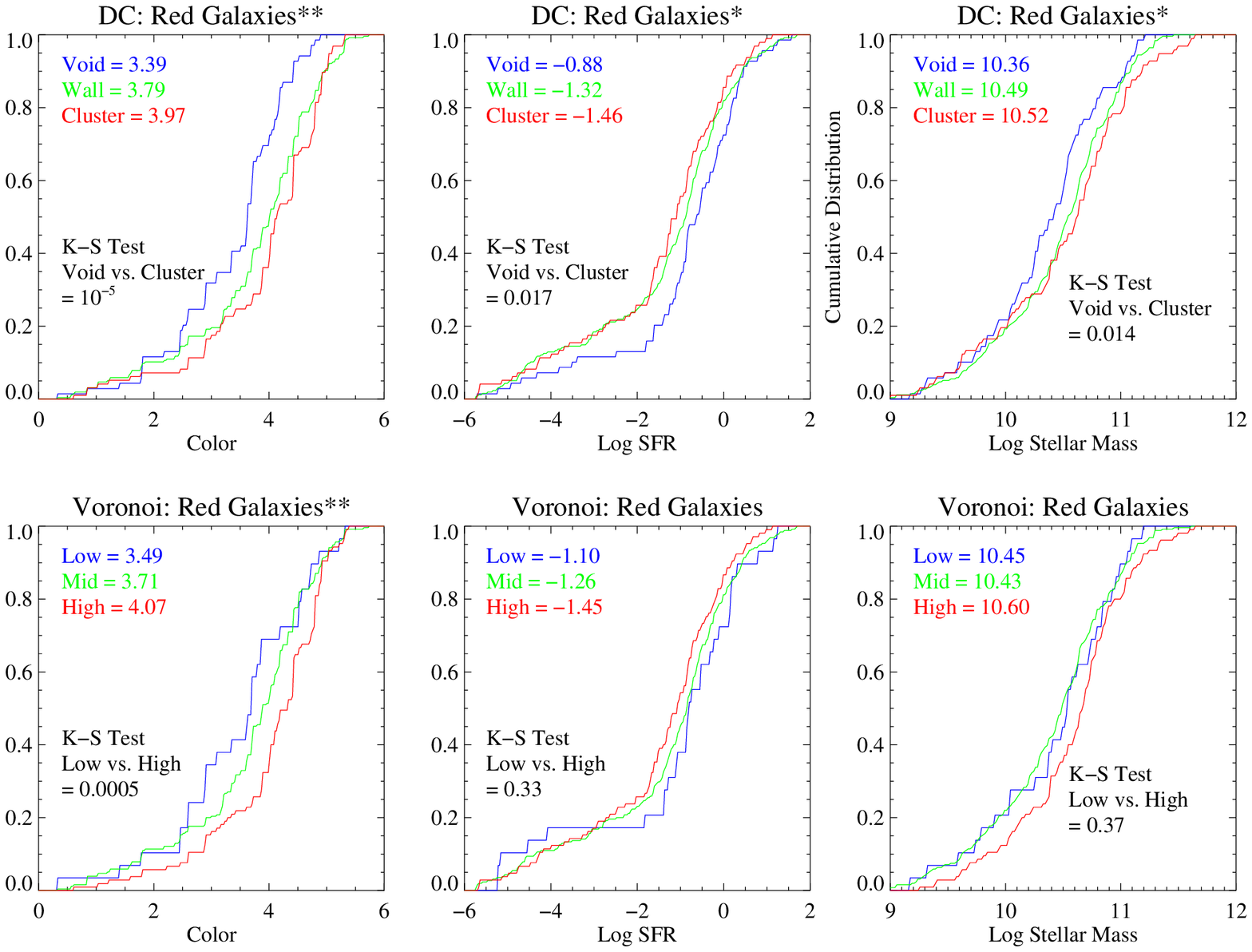}
\caption{The cumulative distributions of color (left), SFR (middle) and stellar mass(right) 
in three environmental bins defined using the DC measure (top row) and Voronoi tessellation (bottom row). 
While both the network and Voronoi measures show that galaxy color is correlated with local density, 
the K-S test values (quoted in the panel legend) suggests that the Void and Cluster populations are better separated by the DC measure. 
The asterisks marks used in Table 1 are marked on the titles. 
}\label{fig:localred}
\end{figure*}

\begin{figure*}[t]
\centering
\includegraphics[height=4.8 in]{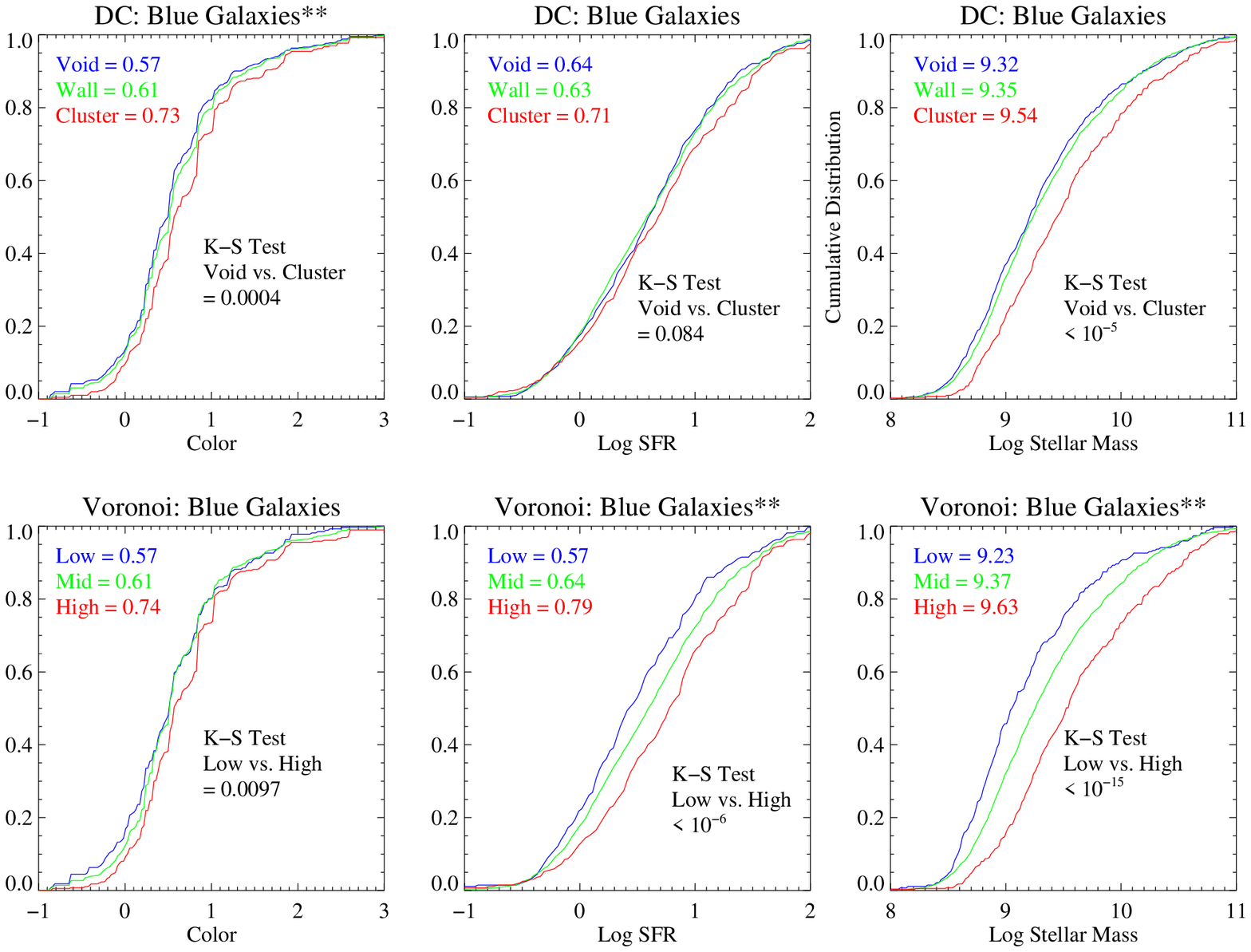}
\caption{The same with Figure~\ref{fig:localred} for blue galaxies.  
In contrast to the results shown in Figure 9 for the red galaxies, 
the blue galaxies are better separated by the Voronoi density measures than by the DC measure.
The stellar mass distribution of blue galaxies shows the small K--S test value, $<10^{-5}$. 
But we do not mark this relation with the double asterisks, since the Wall and Void are poorly separated. 
This poor statistical separation between the Void and Wall for blue galaxies can be found 
on all the top panels of the DC environment. 
}\label{fig:localblue}
\end{figure*}

Using the three network measures we have defined 8 topological classes of galaxies: 
Void, Wall, and Cluster by DC; 
Main Branch and Dangling Leaf by BC; and Kernel, Backbone, and Fracture by CL. 
Since the BC and CL selections are newly introduced by network analysis, 
the comparisons between these classes and the more conventional DC selection 
can help us investigate whether network-based topology can reveal new characteristics of cosmic structures. 
Figure~\ref{fig:compdensity} summarizes the spatial distributions of the selected populations presented 
in Figure~\ref{fig:dgcosmos} -- \ref{fig:clcosmos}, visualizing their different topological selections. 
The green color contours show the distribution of galaxy density at $0.91 < z < 0.94$, 
obtained by the Voronoi-Delaunay method (for details, see Marinoni et al. 2002, Gerke et al. 2005, and Cooper et al. 2005). 
Since the method assigns a single Voronoi-Delauney polygon to each galaxy 
(with the exception of galaxies near the survey edge, which have unbounded Voronoi polygons), the inverse of the polygon volume (or area) 
provides an excellent density measure to each galaxy.  
The contour scale is logarithmic ($\log(\rho/\bar{\rho})$, where $\bar{\rho}$ 
is the average value of the density field). 
To compare this Voronoi tessellation density 
(hereafter, simply called Voronoi density) with our DC measurement, 
we define ``Voronoi High'', ``Voronoi  Middle'', and ``Voronoi Low'' by ranking galaxies according to their Voronoi densities, 
matching the number of galaxies in Cluster, Wall, and Void. 

For each topological class of galaxies, we measure the means and standard deviations 
of various galactic properties (specifically color, stellar mass, 
and star formation rate) from the COSMOS catalog (Capak et al. 2007, McCracken et al. 2012, and Ilbert et al. 2013).
The results are presented in Table 1 (astroph only, at the end of this paper). 
We divide each topological class into two sub-populations, ``red'' and ``blue'' galaxies, 
adopting the two-color selection of Ilbert et al. (2013).
They apply the criteria, NUV $- r >$ 3 $(r - J)$ + 1 and NUV $- r >$ 3.1 in absolute magnitudes, 
to separate ``quiescent'' (``red'' in our terminology) galaxies from ``star forming'' (``blue'') galaxies
, where NUV, $r$ and $J$ are the rest-frame near-UV color (defined in the GALEX NUV 2300\AA), $r$ and $J$ bands.
This color selection can avoid the inclusion of dusty star-forming galaxies with quiescent galaxies and 
can minimize  the uncertainties in k-corrections.
%Due to the large intrinsic variation in galactic properties, 
Due to the uncertainties associated with photometric redshift estimates, 
indirect measurements of SFR, and stellar mass by fitting spectral energy distribution (SED) models, 
the galactic quantities are not clearly distinguished statistically. 
%and they are all covariant with other galactic quantities, such as, dust extinction,   
Hence, to interpret these measurements in Table 1, we focus on two aspects :  
(1) are there any consistent and monotonic trends from low to high environmental selections suggestive of environmental dependencies? and 
(2) are these trends statistically meaningful? 

For example, the color distributions of red galaxies in Cluster, Wall, and Void regions are characterized by mean values of $3.97 \pm1.06$, $3.79 \pm1.14$, and $3.39 \pm0.96$. 
Though the standard deviations are large, the mean values suggest a consistent and monotonic trend of redder colors in denser environments. 
A Kolmogorov--Smirnov (K--S) test suggests that the probability that the colors of the Cluster and Void galaxies are drawn from a common distribution is $10^{-5}$, 
suggesting that their color distributions are statistically different.
%The Kolmogorov--Smirnov (K--S) test result between the Cluster and Void selections is $10^{-5}$. 
%This suggests that the color distributions of Cluster and Void are statistically different. 
This implies that the environment, as defined by the DC measure, affects the colors of red galaxies. 

For the DC, Voronoi, and CL selections in Table 1, we mark the consistent and monotonic trends using ``italic'' fonts and 
tabulate values in ``bold'' fonts when they are most statistically different. Their corresponding cumulative distributions 
and K--S test values are presented through Figure~\ref{fig:localred} -- \ref{fig:clcumul}. 
We identify with double asterisks (**) the relations with the K--S values, $< 10^{-3}$. 
These relations can be considered as reliable environmental effects, given the noisy COSMOS data. 
The single asterisk marks (*) indicate the relations with the K-S values, $< 0.03$. 
In a generous point of view, we can consider them to imply potential environmental effects. 
%From these results, though tantalizing, we find important implications as follows. 
%<< here some discussion about the >> 
%Voronoi density is less successful in the red population than the DC environment.  

\subsubsection{Local Environment : Degree Centrality vs. Voronoi Tessellation Density } 

In this section we compare the results based on our DC measures with those derived using Voronoi tessellation.
Both DC and Voronoi density are local density measurements. 
The difference is that DC uses a fixed linking length, 
while Voroni density is determined by geometric configuration of neighboring galaxies. 
Hence, the scale for the Voronoi density (the size of the Voronoi polygon) varies with location, 
and neighbors outside of the linking length $l_5$ (ignored by our DC measure) can affect the Voronoi density.  

Figure~\ref{fig:localred} shows the color, SFR, and stellar mass  for red galaxies separated using the two environmental measures, DC and Voronoi densities.  The double asterisks marks (**) on the titles represent the relations where the K--S values $<10^{-3}$  (see Table 1). 
For the galaxy colors (left panels), both DC and Voronoi densities show clear statistically significant separations 
implying that the colors of red galaxies are redder in denser environments. 
Since the K--S value is smaller in the DC selection and this trend is also found in the colors of blue galaxies shown in Figure~\ref{fig:localblue}, 
we conclude that \emph{the topological regions selected using the DC measure are a better determinant of galaxy color than the Voronoi-based measures}. 
There are no statistically significant separations for the SFR and stellar mass of red galaxies (the middle and right panels in Figure~\ref{fig:localred}), 
 if we choose the conservative KS significance threshold of $10^{-3}$.
Hence, at this significance threshold,  the SFR and stellar mass of galaxies in this $0.91<z<0.94~$ slice do not appear to depend significantly upon environment, 
as defined by the DC and Voronoi density measures. 
% Hence, in a conservative point of view, neither DC nor Voroni densities shows the environmental effect on the SFR and stellar mass of red galaxies. 
If we lower our threshold of acceptable significance and consider the relations with K-S values $<0.03$ (i.e., those marked by a single asterisk in Table 1), then 
% When accepting the relations of a single asterisk mark (i.e., with the K--S values of $< 0.03$), 
the environmental selection based on DC shows higher significance differences 
in the stellar masses and SFRs in different environments than the selection based 
on the Voronoi density. 
In addition, the three DC-based environmental classes in Figure~\ref{fig:localred} display more consistently monotonic behavior in  SFR and stellar mass, whereas the Voronoi density classes do not; the cumulative lines cross each other, showing no clear differences in different environmental regions. 
This suggests that \emph{the SFR and stellar mass of red galaxies are more likely regulated by the DC environment rather than the Voronoi environment}. 

In contrast to the behavior observed for the red galaxy population, we find that the SFR and stellar mass distributions 
of the blue galaxies are more strongly dependent on the environments measured by the Voronoi density 
than those measured by the DC criterion (see the (**) marked panels in Figure~\ref{fig:localblue}). 
There is no separation in the blue galaxy SFR distributions between any of the DC environmental classes  
(top middle panel in Figure~\ref{fig:localblue})  or the blue galaxy stellar mass distributions 
between the ``Void" and ``Wall" classes (top right panel of Figure~\ref{fig:localblue}).  
In contrast, the Voronoi density environmental classes show clear separations in the SFRs and stellar masses, 
and a monotonic progression of the properties with density.

%On the other hand, Figure~\ref{fig:localblue} show the opposite of the results from Figure~\ref{fig:localred}. 
%The SFR and stellar mass of blue galaxies are more strongly correlated with the Voronoi selection than the DC selection. 
%In the top-middle panel, the SFRs of blue galaxies in Wall are smaller than the Void galaxies, showing the lack of consistency. 
%In the top-right panel, Cluster is well distinct from the others, Void and Wall. But the difference between Void and Wall is vague. 
%This suggests that  \emph{the SFR and stellar mass of blue galaxies are morely likely regulated by the Vornoi environment rather than the DC environment}

In summary:
\begin{enumerate}
\item For red galaxies:
	\begin{itemize}
	\item The galaxy color is a function of environment as measured by both DC and Voronoi density, 
	\item DC may be a better discriminant of galaxy color than the Voronoi density,
	\item SFR and stellar mass are more correlated with DC than with the Voronoi density
	\end{itemize}
\item For blue galaxies:
\begin{itemize}
\item SFR and stellar mass are more correlated with Voronoi density than with DC
\item DC is a poor predictor of blue galaxy properties.
\end{itemize}
\end{enumerate}

To explain the findings, we need to understand the difference between the DC and Voronoi measurement recipes. 
DC is measured using the fixed size of top-hat window; galaxies lying at larger distances than the linking length are ignored. 
In contrast, Voronoi polygons are determined by geometric configurations of neighboring galaxies 
with varying scales. For dense regions, the Voronoi scale can be smaller than the linking length;  
for sparse regions, the Voronoi scale can be larger than the linking length, since the distances from neighboring galaxies 
are more likely larger than the linking length. 
The little  environmental separation between ``Void'' and ``Wall'' in DC measures, in contrast to the success of Voronoi measures, 
also suggests that the contribution of neighboring galaxies outside of the linking length is important for blue galaxies. 
\emph{Therefore, we can characterize the DC environment as ``confined and physical length-dependent locality'', 
while the Voronoi environment as ``versatile and neighbor-dependent locality''. }

``Quenching'' has been suggested as one of  the major factors regulating star formation in red galaxies, while  
``gas accretion'' to regulate the star formation of blue galaxies. This implies that the major mechanism forming 
the DC environment is the quenching process, while the major mechanism forming the Voronoi environment is the gas accretion. 
The quenching process is a localized ``inside-out'' process mostly contributed by quasar and stellar feedback, 
while the gas accretion is an ``outside-in'' gas flow more depending on the overall gas distribution in larger local scales.  
This can explain the DC and Voronoi environmental effects, resulting in the implication that 
the quenching process in red galaxies is a scale-confined local phenomenon less dependent on neighboring galaxies, 
while the gas accretion to blue galaxies is a more interactive and extended phenomenon depending more on the configuration of neighboring galaxies.

\subsubsection{Topological Environment : Closeness Centrality}

\begin{figure*}[t]
\centering
\includegraphics[height=4.8 in]{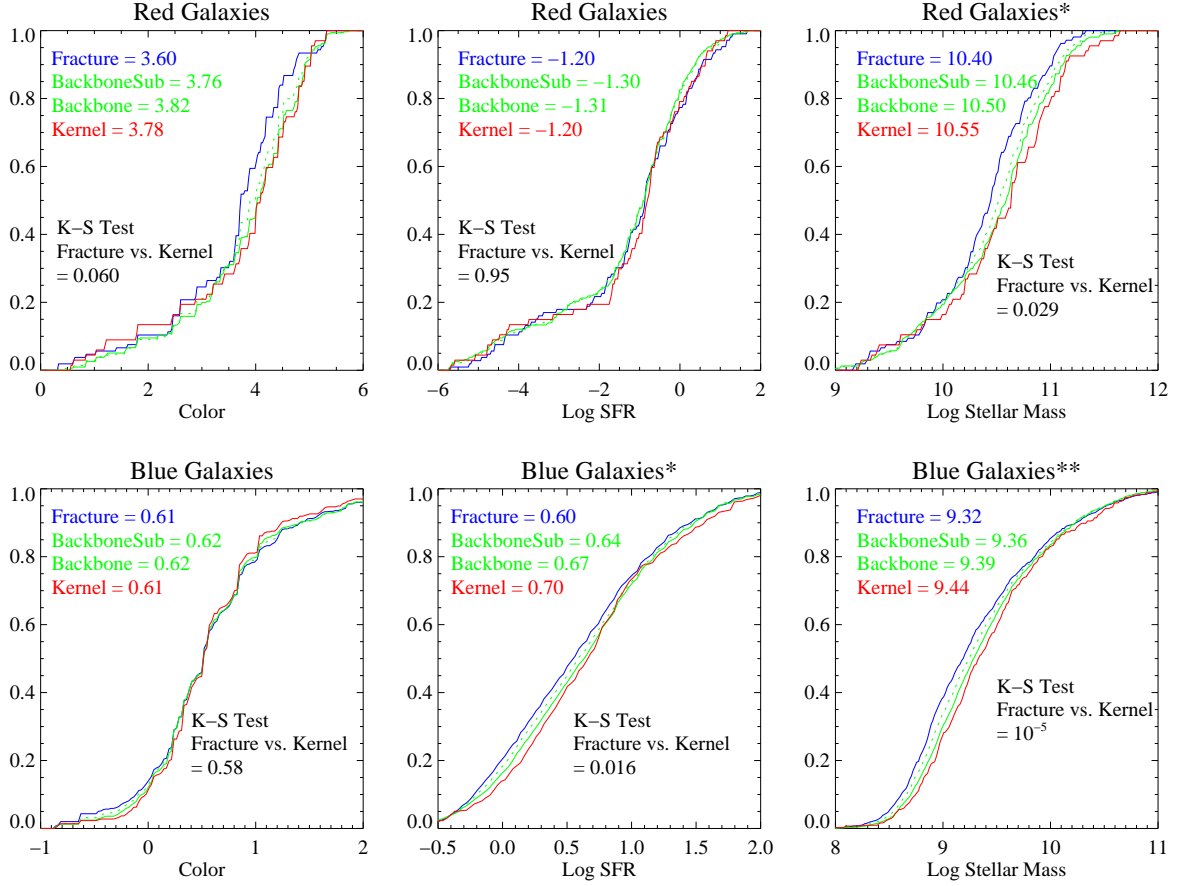}
\caption{The cumulative distributions and K--S test values for the CL selection; Fracture (blue), BackboneSub(green-dotted), Backbone(green), and Kernel(red).  
%The  SFR of red galaxies (top-middle) and the color of blue galaxies (bottom-left) show the collapsed distributions 
%for Fracture, BackboneSub, Backbone, and Kernel. 
The SFR distributions of the red galaxies (top-middle) and the color distributions of the blue galaxies (bottom-right) appear to show no variation with CL measure. 
%This implies that those two quantities are smeared out to the cosmic averages in the scales of the CL selection. 
The colors of red galaxies (top-left) show somewhat different distributions but their trends have neither consistency nor monotonic behavior, 
%Their averages values are fluctuating near the sample mean value, 3.77, 
implying the lack of CL dependency. 
However, the other three panels marked with the single and double asterisks on the titles 
show the relatively consistent and monotonic behaviors with the K--S test results of 0.016, 0.029, and $10^{-5}$. 
%These trends are more interesting when considering the collapsed featureless properties are also derived from the same collections of galaxies.  
}\label{fig:clcumul}
\end{figure*}

\begin{figure*}[t]
\centering
\includegraphics[height=4.8 in]{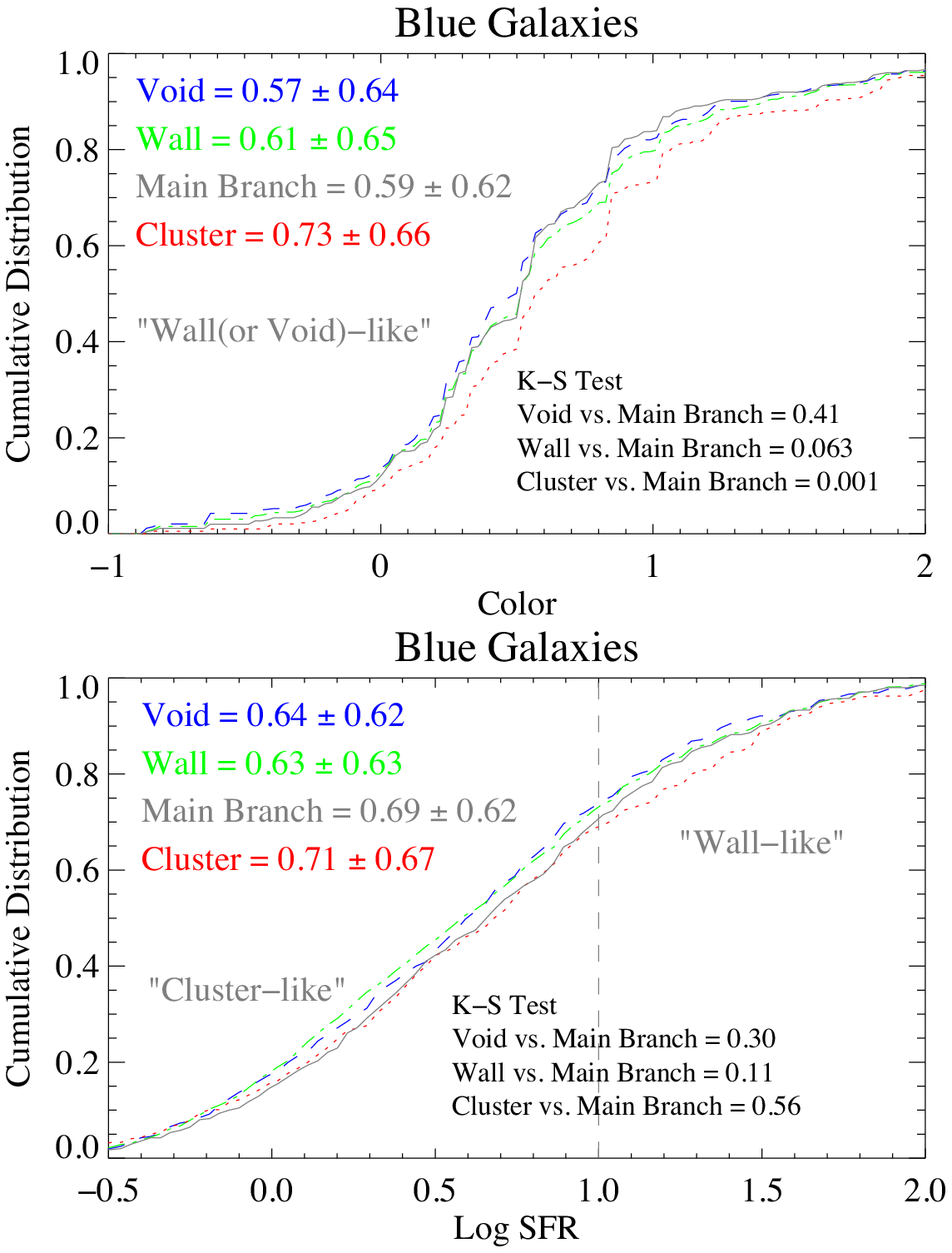}
\caption{The cumulative distributions of colors and SFRs of blue galaxies for Void, Wall, Cluster, and Main Branch, and 
related K--S test results. 
The color of Main Branch is statistically different from Cluster (bluer than Cluster), showing the ``Wall (or Void)-like'' behavior, 
while the SFR of Main Branch is more ``Cluster-like'', especially for SFR $< 10 ~\text{M}_\odot \text{yr}^{-1}$;  
near SFR $\approx 10 ~\text{M}_\odot \text{yr}^{-1}$, the SFR of Main Branch seems to change from being ``Cluster-like'' to being  ``Wall (or Void)-like''. 
This transition possibly implies that the star formation over $\approx 10 ~\text{M}_\odot \text{yr}^{-1}$ needs an additional boost by Cluster-like environment; 
i.e., wet mergers in the local DC environment. Overall, Main Branch seems to be an intermediate phase between Wall and Cluster. 
}\label{fig:mbcumul}
\end{figure*}

%Now we investigate the CL environment, which turns out to be connected with the above implication. 
The DC and Voronoi density are measures of local density; in contrast, the CL and BC measures depend on the entire structure of network.  
Hence, the CL and BC measures reflect the more global environment and its topological structure.
%Therefore, the CL and BC environments are global and/or topological. 

The highest CL vertex is located at the topological center (CL center) of the network. 
Its nearby vertices are generally next highest CL vertices. 
%Hence, a selection of high CL vertices are clustered near the CL center in a bulk size. 
%This bulk keeps growing and eventually fills out Backbone, as we keep including next highest CL vertices. 
Hence,  selecting the highest and next highest CL vertices identifies connected clustered regions, eventually filling out the ``Backbone" of the structure.
Due to this property, the measured CL values gradually vary throughout the scale of system size; 
across the 1 degree ($\approx 54$ Mpc) in our COSMOS network. 

The difference between the CL and DC environments can be described figuratively by  
the difference between a suburb area in a large city such as Los Angeles (LA) and a central urban area in a small city such as Tucson. 
Since the DC environment is defined by a local window, the central urban area in Tucson has a high DC value. 
However, since LA is the largest city in the west coast of the United States, 
the highest CL vertex is located in LA  and the suburb areas of LA have higher CL values than the central urban area in Tucson, 
despite having lower local densities than the Tucson's urban area.  
% Due to this difference, the CL environment  represent a global scale environment based on network topology. 

In our network, the Kernel region is composed of (almost) all galaxies within $0.2$ degree diameter ($\approx 11$ Mpc)  
and Fracture over 0.4 degree ($\approx 22$ Mpc) throughout the 1 degree ($\approx 54$ Mpc) survey area 
as shown in Figure~\ref{fig:compdensity}. 
These scales are large enough to smear out any variation in the galactic properties to cosmic averages. 
Indeed, Table 1 and Figure~\ref{fig:clcumul} demonstrate that  
the distributions of SFRs of red galaxies (top-middle) and colors of blue galaxies (bottom-left) 
are nearly identical for Fracture, BackboneSub, Backbone, and Kernel, 
implying that the properties are ironed out to the averages on these selection scales.  

Interestingly, the other four panels do show differences. 
The stellar masses of blue galaxies (bottom-right panel) show statistically reliable  
separations in the CL selection, implying the gradient of stellar mass 
of blue galaxies across the 1 degree scale ($\approx 54$ Mpc). 
The stellar mass of red galaxies (top-right panel; especially for $> 10^{10} ~\text{M}_\odot$) 
and the SFRs of blue galaxies (bottom-middle panel; especially for $ < 10 ~\text{M}_\odot \text{yr}^{-1}$) 
also show a possible dependence on the CL environmental measure. 

The colors of red galaxies (top-left panel) show a relatively low K--S value, 0.06, 
%but the cumulative lines cross each other showing lack of consistency and 
but the cumulative distributions do not show monotonic variations with the environmental classes. 
The average colors are fluctuating near the sample mean value, 3.77, 
implying the lack of CL dependency like the two ironed-out quantities, the colors of blue galaxies and the SFRs of red galaxies; 
but the fluctuation is larger.  

%About the CL environment, we can summarize the implications from Figure~\ref{fig:clcumul} as : 
% To summarize the implications about the CL environment from Figure~\ref{fig:clcumul},  
In summary:
\begin{enumerate}
\item For red galaxies:
	\begin{itemize}
	\item There is no dependence of color and SFR on CL environment, 
	\item There is some evidence for a dependence of stellar mass on CL environment.
	\end{itemize}
\item For blue galaxies:
\begin{itemize}
\item There is no dependence of color on CL environment, 
\item There is some evidence for a dependence of SFR on CL environment, 
\item There is a stronger evidence for a dependence of stellar mass on CL environment. 
\end{itemize}
\end{enumerate}

In \S~3.2.1,  we showed that 
the colors, SFRs, and stellar masses of red galaxies and the color of blue galaxies are more likely shaped by the DC environment, 
while  the SFRs and stellar masses of blue galaxies by the Voronoi environment. 
When comparing the CL results with these DC and Voronoi results, we can find three interesting results: 
(1) The colors of red and blue galaxies and the SFR of red galaxies, which show a dependence on the DC environment, 
do not show any dependence on the CL environment; we refer to this as the ``exclusive CL-DC connection''. 
(2) In contrast, the SFR and stellar mass of blue galaxies show both a dependence on the CL and Voronoi environments 
(i.e., the ``inclusive CL-Voronoi connection''). 
(3) Finally, the stellar mass of red galaxies shows dependencies both on the CL and DC environments. We discuss these in turn.

%(1) The three quantities showing no CL dependence, the colors of red and blue galaxies and the SFR of red galaxies, 
%are shaped by the DC environment; \emph{the ``exclusive CL--DC connection''}.   
%(2) Conversely, the two quantities showing CL dependencies, the SFR and stellar mass of blue galaxies, 
%are shaped by the Voronoi environment; \emph{the ``inclusive CL--Voronoi connection''}. 
%(3) The stellar mass of red galaxies, which is more likely shaped by the DC environment, 
%but also shows the CL dependence. 
%We discuss these in turn.

The exclusive CL--DC connection   might imply that the quenching process is also independent on its global position in the CL scale.  Since the DC environment is more localized and less dependent on neighbors,  the exclusive connection suggests a more local behavior of the quenching process.  The inclusive CL--Voronoi connection indicates that the SFR and stellar mass of blue galaxies depend not only on the local environment  but also on the more global topology (as sampled by the CL measure). Since the Voronoi selection still shows the better correlation than the CL selection, the SFRs and stellar masses of blue galaxies are more affected by their local neighbors. 
However, the CL dependence reflects that the global (and topological) positions are also important to shape the SFR and stellar mass of blue galaxies. 
This implies that the global shapes of the Universe from Fracture, BackBone, to Kernel, affect  to regulate the SFR and stellar mass of blue galaxies. 
{\it In other words, the fueling of star formation in blue galaxies is dependent not only on the local environment but also on the larger-scale environment. This ``global and topological'' dependence of gas accretion onto blue galaxies is in contrast with the purely local dependence of quenching processes.}
%\emph{This ``global and topological'' dependence of fueling gas to blue galaxies 
%shows an excellent contrast with the strong ``locality'' of the quenching process; } 
%{\bf i.e., the extended ``Voronoi-CL environment'' for blue galaxies vs. the local ``DC environment'' for red galaxies.} 

 The third result suggests that while the major mechanism which shapes the stellar mass distributions of red galaxies 
is still the local quenching process (based on the DC effect being stronger than the CL effect; see Table 1), 
% like the other three quantities, the SFR of red galaxies and colors of blue and red galaxies.
%However, the exclusive CL dependence of the stellar mass of red galaxies suggests that 
%\emph{the stellar mass evolution of red galaxies can not be explained by the local quenching only.} 
% However, the fact that the stellar mass distributions do show a statistically meaningful dependence 
% on the CL environmental measures suggests that 
the local effects are not solely responsible for shaping these distributions. 
When comparing the galaxies selected by the Kernel and Cluster selections, 
we find that red galaxies in the Kernel show average (i.e., cosmic mean) values of SFR and color, 
where as their counterparts in the Cluster selection show traits characteristic of dense environments. 
{\it Surprisingly, the stellar mass distributions are similar in both the Kernel and Cluster subsets, 
despite having different spatial distributions. This might imply that there is a global (topological) 
channel for red galaxy growth, analogous to the CL-dependence of gas fueling for blue galaxies. } 

We speculate that this CL dependence of stellar mass for red galaxies arises 
because the bulk of stellar mass in red galaxies is assembled in an early phase through gas accretion and star formation 
(i.e., during which they would appear as blue galaxies). 
%This legacy of ``once it used to be blue'', {\bf when stellar mass grows by the extended Voronoi--CL environment, can be encrypted 
%as the CL dependence of stellar mass in red galaxies, 
%since stellar mass is a cumulative quantity integrated through the entire history of star formation and gas accretion. 
%Unlike this cumulative (or integral in time) property of stellar mass, 
%star forming activity and color of red galaxies are more transient (or differential in time) phenomena, mostly affected by the local quenching process. 
%Therefore, this integral feature of stellar mass can explain 
%why the stellar mass of red galaxies can show the CL dependence, which is more typical for blue galaxies.}
The CL dependence of stellar mass in red galaxies therefore arises during this growth phase, 
during which the galaxies likely appear as blue galaxies. 
As mentioned above, the SFR of blue galaxies do indeed show a dependence on the CL environment, 
in support of this picture. 
In contrast to the growth phase, the quenching of star formation 
in these systems (and therefore their transformation from blue to red galaxies) appears 
to be a more local phenomenon, resulting in the more local dependence on environment of the color and SFR of red galaxies.

\emph{To summarize, the colors of red and blue galaxies and the SFRs of red galaxies are likely shaped by the DC environment 
related to local quenching process. The SFR and stellar mass of blue galaxies are likely shaped by the Voronoi environment 
and extended to more global scale of the CL environment (the Voronoi-CL environment). 
The stellar mass of red galaxies seems to be shaped by both of the DC and CL 
environments suggesting the two-phase stellar mass growths; blue phase in the Voronoi-CL environment and red phase in the DC environment. }

\subsubsection{Topological Environment : Betweenness Centrality}

{Unlike the other density measures, DC, CL, and Voronoi density, the selection by BC identifies filamentary density enhancements: 
the Main Branch traces galaxies on filamentary structures and the Dangling Leaf traces 
galaxies lying at the outer envelopes of structures or in voids. 
Though the selections are unique, the COSMOS photo-z dataset 
do not show any significant trends in galaxy color, SFR or stellar mass in this topological classification. It is possible that a 
larger spectroscopic sample with better data would allow the BC selection to be used with more discriminatory power.
% When a good spectroscopic sample is available, the following arguments can be better investigated. 

When comparing Main Branch with the DC selection, Void, Wall, and Cluster in Table 1, 
the red galaxies in Main Branch show similar SFRs and stellar masses with the red galaxies in Cluster, but bluer colors than in Cluster. 
This implies that the red galaxies in filaments already have suppressed SFRs and massive stellar content like the red galaxies in Cluster 
(note that our red shift slice is 0.91 - 0.94), but still exhibit slightly bluer colors than the Cluster's red galaxies. 
This possibly implies that a higher fraction of ``Green Valley'' galaxies reside in filaments rather than in cluster regions at $z\approx 0.9$.

While the red galaxies in the Main Branch are ``Cluster-like'' except for their bluer colors, 
the blue galaxies in the Main Branch are more ``Wall(or Void)-like''. 
Figure~\ref{fig:mbcumul} shows the cumulative distributions of colors and SFRs for Void, Wall, Main Branch, and Cluster for blue galaxies. 
The colors of blue galaxies in Main Branch show a clear difference from Cluster 
through the whole color range from -1 to 2 with the K--S value, 0.001. Therefore, at least for colors, 
Main branch blue galaxies are ``Wall(or Void)-like''. 
On the other hand, the SFRs show a more interesting intermediate behavior than the colors (the bottom panel of Figure~\ref{fig:mbcumul}). 
The mean SFR of Main Branch, 0.69, is closer to the SFR of Cluster, 0.71, than the Wall's SFR, 0.63. 
The cumulative distribution starts with ``Cluster-like'' behavior and continues the behavior up to SFR$< 10 $~M$_\odot $ yr$^{-1}$ 
(the red-dotted line and grey-solid line). 
For SFR$> 10 $~M$_\odot $ yr$^{-1}$, the cumulative distribution deviates from its ``Cluster-like'' behavior and becomes more ``Wall-like''. 
This transition possibly implies that the star formation over $\approx 10 ~\text{M}_\odot \text{yr}^{-1}$ needs an additional boost by Cluster-like environment; 
i.e., wet mergers in the DC environment. 
Main Branch galaxies appear to be intermediate in their star-formation properties between Wall and Cluster populations.

%Dangling Leaf shows more marginal statistical results with the current data set. 
%But the implication from Dangling Leaf is also interesting. 

The spatial distributions of galaxies in the Dangling Leaf and Void selections are similar (see Figure~\ref{fig:compdensity}). The mean SFR of Dangling Leaf blue galaxies resembles that of the Fracture population (see Table 1). Dangling Leaf red galaxies exhibit the highest mean SFR of all topological selections (perhaps because quenching processes are inefficient in these regions). Since the SFRs of blue Dangling Leaf galaxies are among the lowest of all topological classes, this environmental region has the smallest difference in SFRs between red and blue populations. Since the Dangling Leaf samples galaxies lying at the outer boundary of the cosmic mass distribution, this might imply that both accretion and quenching processes are inefficient at these edges. 

\section{Future Improvements}\label{sec:future}

%\subsection{Adapting and Improving Network Tools for Astronomy}

%What we have presented in this paper is a small part of possible network application. 
%Though we have only shown the DC measure for local density, we have also measured  
%Katz centrality, PageRank, and eigenvector centrality (variants of DC; described in the following subsection), 
%which turn out not applicable to our comic networks. 
%We also defined the weighted DC (wDC) to find a better BC measure to identify filamentary structures. 
%Therefore, basically, there is no limit to define new network measurements or new recipes to build networks. 
%In this section, we discuss some directions for new network quantities and network recipes  in the future studies. 

In this paper, we have investigated some simple applications of network analyses
to understanding the topology of cosmic structure and its relationship to galaxy properties. 
Many other complex measures are possible and may better quantify the cosmic network 
and perhaps improve our understanding of how galaxy properties correlate with the topology 
in their local environment. Here, we discuss some possible directions for future research.

%\subsubsection{Customized Network Measures for Astronomy}
\subsection{Customized Network Measures for Astronomy}

%We have used the graph library, \emph{igraph}, to calculate network measures. 
Except for the weighted DC measure (wDC), all quantities we have presented here are 
common network measures used in complex networks. 
The DC and Voronoi results suggest that we may invent  
new centralities to trace local density environments for better correlations with gas inflows and quenching processes.

One customizable centrality, $x_i$, in a general form, can be defined as 
\begin{equation}\label{eq:newcent}
x_i = \alpha \sum\limits_{j}A_{ij} x_j + \beta \sum\limits_{j}A_{ij} w_j + \gamma v_i, 
\end{equation}
where, $\alpha$, $\beta$, and $\gamma$, are customizable constants, $A_{ij}$ the adjacency matrix, 
$w_j$ customizable scalar weights coupled with the adjacency matrix, and $v_i$ scalar weights uncoupled the adjacency matrix. 
This linear equation is a generalized version of Katz centrality (Katz 1953 and Newman 2010), which can cover 
most variants of DC used in complex networks. 
When $\alpha=0$, $\beta=1$, $w_j \equiv 1$, and $\gamma=0$, Equation~\ref{eq:newcent} represents DC. 
When $\alpha= \lambda_1^{-1}$, $\beta=0$, and $\gamma=0$, where $\lambda_1$ is the largest eigenvalue of $A_{ij}$, 
Equation~\ref{eq:newcent} represents ``eigenvector centrality''.  
When $\beta=0$ and $\gamma=1$, Equation~\ref{eq:newcent} represents  Katz centrality with weights of $v_i$. 
%Hence, Equation~\ref{eq:newcent} is a quite generalized form to cover most DC variants. 

To reduce free parameters for a more practical centrality in Equation~\ref{eq:newcent}, we set $w_j \equiv 1$.  
Then, we obtain  
\begin{equation}\label{eq:newcentfinal}
x_i = \alpha \sum\limits_{j}A_{ij} x_j + \beta k_i + \gamma v_i, 
\end{equation}
where $k_i$ is a DC for the vertex $i$ derived by $w_j \equiv 1$. 
To count the voronoi density contribution, we may define a ``Voronoi Weight'' as the ratio of total survey volume 
to each Voronoi polyhedron (or polygon) as 
\begin{equation}
v_i = \frac{\text{total survey volume}}{\text{volume of each Voronoi polyhedron}}, 
\end{equation}
and use these for the uncoupled scalar weights in Equation~\ref{eq:newcent}. 
 
This is a Katz centrality with parametrized weights of $\beta k_i + \gamma v_i$. 
By controlling the three parameters, $\alpha$, $\beta$, and $\gamma$, 
%though the computation can be demanding to cover this three dimensional parameter space, 
we can find better centrality measures to represent local cosmic density. 
%PageRank used in the search engine, \emph{Google}, is a variant of this Katz centrality. 
%Its empirical choice of $\alpha$ is 0.85 for the optimal ranks of web documents. 
PageRank, used by the search engine {\it Google} to determine the optimal ranks of web documents, 
is a variant of the Katz centrality with the empirical choice of $\alpha=0.85$ and constant weight of $\beta k_i + \gamma v_i \equiv 1$; 
more specifically, the definition of PageRank is slightly different from Katz centrality because WWW is a directed network (see, Page et al. 1999). 
 Future studies may find the optimal set of $\alpha$, $\beta$, and $\gamma$ for cosmic local environments 
when investigating large samples of galaxies with spectroscopic redshifts or new suites of sophisticated cosmological simulations.

%\subsubsection{Nonparametric Recipes to Build Networks}
\subsection{Nonparametric Recipes to Build Networks}

The network recipe used in this paper depends on linking length. Shape finders based on the Hessian matrix also depend on smoothing length. 
These parametric representations of cosmic matter distribution are necessary if there is a physical reason for the scale. 
For example, since there is a physical length in the two-point correlation function of dark matters 
from one halo to two halo contributions ($\approx 2 h^{-1} $ Mpc at $z=0$ in the MS2; Boylan-Kolchin et al. 2009), 
at least for halo networks, this emergent scale length needs to be considered to define neighbors. 

On the other hand, when there is no physical reason (or constraint) for the scale, 
the parameter is only an unnecessary and artificial construct. 
For shape finders based on the Hessian matrix, 
Cautun et al. (2013) introduced their new NEXUS algorithm to remove the unnecessary scale dependence. 
They try multiple scales of smoothing and find consistent structures independent of the scales. 
For network representations, we have a good conventional example of self-consistent network. 
The Voronoi--Delaunay meshes (or complexes) are nonparametric  
structures self-consistently derived from a given population (e.g., Marinoni et al. 2002 and Gerke et al. 2004). 
If we connect all first Delaunay neighbors, we can obtain a unique nonparametric (self-consistent) network; 
one may call this network ``Delaunay Network''. 
This shows the possibility that we can find a useful self-consistent network recipe in the future studies. 

%(((Maybe, we need more discussions about this parameter dependency. )))
%Many clustering studies and the observed sizes of voids and filaments show 
%that there are physical scales to define ``neighbors'' in galaxy (or halo) distributions  
%(i.e., not scale-invariant fractals). We can not call the galaxies ``neighbors'', 
%which are hundreds mega parsecs away from the interested galaxy.  

%Figure blah shows the BC and CL selections for L4 and $l_6$. 
%Definitely L10 shows a different 

%Hence, the scale lengths to define neighbors are not totally arbitrary. 
%Self-consistent networks need to be within the physically acceptable range. 
%Overall, we need to develop both parametric and nonparametric recipes for specific target populations.  

\section{Summary}

%The purpose of this paper is to demonstrate how the tools of complex networks 
%can be utilized to investigate the structures of the Universe.  
In this paper we have attempted to demonstrate that the analyses tools developed 
to analyze complex networks can be applied to the investigation of cosmic structures 
and can potentially provide useful insights into the relationship between 
the internal properties of galaxies and their topological environment.
We have presented the basics of network theory and 
described simple recipes to define and measure the cosmic network. 
Selecting galaxies at $0.91<z<0.94$ from the COSMOS catalog, 
we constructed a network using a simple cylindrical top- hat window, 
calculated three centrality measures (DC, CL, BC), and defined 8 (overlapping) 
topological classes of galaxies (i.e., DC: Void, Wall, Cluster; BC: Main Branch and Dangling Leaf; and CL: Kernel, Backbone and Fracture). 
We then investigated the existence of any relationships between these topological classes and galaxy properties 
(colors, stellar masses and star formation rates). Finally, we compared any correlations with those measured 
using the more traditional Voronoi-tessellation-based density measures.

The two local density measures, DC and Voronoi density, show intriguing ``environment -- population'' connections : 
in particular, at $z\sim0.9$, we find that the red galaxy population properties are better correlated 
with topology defined using DC measures, whereas the blue galaxy population properties 
are better correlated with the Voronoi density. We speculate that this difference suggests 
that the main mechanisms shaping the galaxy properties 
(say, quenching and gas fueling in the case of red and blue populations respectively) 
may be traced by different measures.
In the discussion section, we propose a new parametrized Katz centrality 
as a new network measure for local cosmic environment. 
From the CL measurement, we have found non-local dependencies of galactic parameters, 
the most significant being the stellar mass of blue galaxies. 
The stellar mass of red galaxies and the SFR of blue galaxies also show some dependence on the CL environmental measure.
Since the scales of the CL selection are large enough to smear out most of galactic properties to cosmic averages, 
these CL environmental effects are very interesting. 
Finally, we find possible correlations with BC environment: 
Main Branch galaxies appear to be intermediate in their SFR and color of blue galaxies between Cluster and Wall (or Void). 
Dangling Leaf galaxies show the smallest gap between the SFRs of blue and red galaxies. 

In this paper we analyzed a galaxy sample selected on the basis of photometric redshift. 
The resulting large positional uncertainty and inability to resolve three-dimensional topological 
structures with accuracy undoubtedly results in washing out any underlying correlations 
between galaxy properties and topology. Despite this, the results presented here 
are suggestive of trends in galaxy properties that depend on the topology of the local environment. 
Future studies that (a) construct better topological measures than the simple ones described here and 
(b) apply them to large samples of galaxies with spectroscopic redshifts, may be better able 
to investigate these dependencies. In particular, applying the analyses 
in parallel to the new suite of sophisticated cosmological simulations 
(e.g., Springel et al. 2010, MNRAS, 401, 791; Vogelsberger et al. 2014) 
will help elucidate the driving forces between these topological correlations.

\acknowledgments
We thank Nick Scoville and Olivier Ilbert for generously making available the photometric redshift catalog 
for the COSMOS field and for their advice. We are also grateful to Bruno Coutinho, Albert-L\'aszl\'o Barab\'asi and 
Miguel Calvo for their advice and assistance with network analyses and for comments on the manuscript. 
The Millennium-II Simulation databases used in this paper and the web application providing online access 
to them were constructed as part of the activities of the German Astrophysical Virtual Observatory (GAVO).
SH and AD's research activities are supported by the National Optical Astronomy Observatory (NOAO). 
AD thanks the Radcliffe Institute for Advanced Study and the Institute for Theory and Computation 
at Harvard University for their generous support during the period when this paper was written. 
NOAO is operated by the Association of Universities for Research in Astronomy (AURA) under 
cooperative agreement with the National Science Foundation. 
SH's research was also supported by NASA through grants JPL\#1497290 and HST GO11663.

\clearpage

%\begin{landscape}
\begin{turnpage}
\begin{deluxetable}{crrrrrrrrrrrrrrrr}
\tabletypesize{\scriptsize}
%\rotate
\tablecolumns{13}
\tablewidth{0pc}
\tablecaption{Topological Selections and their Galactic Properties of Blue and Red Populations}
\tablehead{
\colhead{Selections\tablenotemark{a}}& \colhead{Total\tablenotemark{b}} & \multicolumn{2}{c}{Fraction~\tablenotemark{c}} & \colhead{} & \multicolumn{2}{c}{Color~\tablenotemark{d}} 
&\colhead{} & \multicolumn{2}{c}{Log SFR~\tablenotemark{d}}  & \colhead{} &\multicolumn{2}{c}{Log Stellar Mass~\tablenotemark{d}}  \\
\cline{3-4}  \cline{6-7} \cline{9-10} \cline{12-13} \\
\colhead{}& \colhead{} & \colhead{Red} & \colhead{Blue} &\colhead{} & \colhead{Red} & \colhead{Blue}
&\colhead{} & \colhead{Red} & \colhead{Blue} &\colhead{} & \colhead{Red} & \colhead{Blue} 
}
\startdata
All &  3366 &  0.125   &   0.875  &  &  3.77 $\pm$1.11&  0.62 $\pm$0.65&  &  -1.28 $\pm$1.68&   0.65 $\pm$0.64&  &  10.48 $\pm$0.54&  9.37 $\pm$0.60    \\ 
\hline \hline \\
Cluster & 492 & 0.197  &  0.803 &  & {(**) \bf{\emph {3.97 $\pm$1.06}}}\tablenotemark{e}&  {(**)\bf{\emph { 0.73 $\pm$0.66}}}&  & (*) {\bf{\emph {  -1.46 $\pm$1.66}}}&  0.71 $\pm$0.67&  & (*) {\bf{\emph {  10.52 $\pm$0.59}}}&  {\it  9.54 $\pm$0.62}\\ 
Wall  &  2120  &   0.120  &   0.880  &  & {(**)\bf{\emph {  3.79 $\pm$1.14}}}\tablenotemark{e}&  {(**)\bf{\emph {  0.61 $\pm$0.65}}}&  &  (*) {\bf{\emph {  -1.32 $\pm$1.71}}}&   0.63 $\pm$0.63&  &(*)  {\bf{\emph {   10.49 $\pm$0.52}}}& {\it   9.35 $\pm$0.59} \\ 
Void &  754  &   0.092  &   0.908 &  & {(**)\bf{\emph {  3.39 $\pm$0.96}}}\tablenotemark{e}& {(**)\bf{\emph {  0.57 $\pm$0.64}}}&  &(*)  {\bf{\emph {   -0.88 $\pm$1.53}}}&   0.64 $\pm$0.62&  & (*)  {\bf{\emph {  10.36 $\pm$0.50}}}& {\it   9.32 $\pm$0.60}  \\ 
\cline{1-13}\\
Voronoi High & 492 & 0.213  &  0.787 &  & (**) {\it 4.07 $\pm$0.98}& (*)  {\it 0.74 $\pm$0.67} &  &  {\it -1.45 $\pm$1.60}& {(**) \bf{\emph { 0.79 $\pm$0.65}}}&  &  10.60 $\pm$0.50&{(**) \bf{\emph { 9.63 $\pm$0.62}}}\\ 
Voronoi Middle  &  2120  &   0.120  &   0.880  &  &(**)  {\it 3.71 $\pm$1.12}& (*)  {\it 0.61 $\pm$0.63}&  & {\it  -1.26 $\pm$1.69}&  {(**) \bf{\emph { 0.64 $\pm$0.63}}}&  &   10.43 $\pm$0.55& {(**) \bf{\emph { 9.37 $\pm$0.59}}} \\ 
Voronoi Low &  754  &   0.081  &   0.919 &  &(**)  {\it 3.49 $\pm$1.14} &(*)  {\it 0.57 $\pm$0.68} &  &  {\it -1.10 $\pm$1.76}&  {(**) \bf{\emph { 0.57 $\pm$0.62}}}&  &   10.45 $\pm$0.53 & {(**) \bf{\emph { 9.23 $\pm$0.58}}}  \\ 
\hline \hline \\
Kernel &  500 & 0.134 & 0.866&  &3.78  $\pm$1.23&   0.61  $\pm$0.60&  &   -1.20  $\pm$1.72& (*)  {\it 0.70  $\pm$0.63}&  & (*)  {\it 10.55  $\pm$0.57} &  (**) {\it 9.44  $\pm$0.60} \\ 
Backbone &  2311 & 0.136 & 0.864&  & 3.82  $\pm$1.10&  0.62  $\pm$0.64&  &  -1.31  $\pm$1.67& (*) {\it 0.67  $\pm$0.63}&  & (*) {\it 10.50  $\pm$0.55}& (**) {\it  9.39  $\pm$0.59} \\ 
Fractures &  1055 & 0.100 & 0.900&   & 3.60  $\pm$1.11&  0.61  $\pm$0.67&  &   -1.20  $\pm$1.69&(*) {\it 0.60  $\pm$0.64} &  &(*) {\it 10.40  $\pm$0.49} & (**)  {\it 9.32  $\pm$0.63} \\ 
\cline{1-13}\\
Main Branch  & 500 & 0.102 & 0.898 &  &  3.85  $\pm$1.17&   0.59  $\pm$0.62&  &   -1.47  $\pm$2.02&   0.69  $\pm$0.62&  &   10.55  $\pm$0.46&   9.40  $\pm$0.58 \\ 
Dangling Leaf & 917 & 0.123 & 0.877&   &3.46  $\pm$1.11&   0.60  $\pm$0.66&  &  -0.84  $\pm$1.37& 0.61  $\pm$0.63&  &  10.39  $\pm$0.51&   9.32  $\pm$0.59 \\
\enddata
\tablenotetext{a}{ The galaxy selections by each topological feature.} 
\tablenotetext{b}{ The total number of galaxies for each selection.} 
\tablenotetext{c}{ The fractions of blue and red galaxies for each selection. 
The red galaxies are selected by the criteria, NUV $- ~r > 3 (r - J)$ + 1 and NUV $-~ r >$ 3.1 in absolute magnitudes (Ilbert et al. 2013).} 
\tablenotetext{d}{ The quantities adopted from the COSMOS catalog (Scoville et al. 2013). }
\tablenotetext{e}{For the DC, Voronoi, and CL selections, we mark the consistent and monotonic trends 
using ``italic'' fonts and tabulate values in ``bold'' fonts when they are most statistically different. 
The corresponding cumulative distributions and K--S test values for these trends are presented through Figure~\ref{fig:localred} -- \ref{fig:clcumul}.
We use the double asterisks (**) for the relations with the K-S values, $<10^{-3}$. 
These relations show statistically acceptable separations, though in a conservative view and considering the noisy COSMOS data. 
The single asterisk marks (*) indicate the relations with the K-S values, $<0.03$, which represent possible trends, albeit at lower significance. % In a generous point of view, we can consider them to imply the potential environmental effects. 
The speculative arguments presented in this work can be more clearly investigated in the future spectroscopic surveys. }
\end{deluxetable}
%\end{landscape}
\end{turnpage}

\end{document}